\documentclass[a4paper]{article}
\usepackage{amsmath,amssymb,graphicx}
\usepackage{microtype}
\usepackage{hyperref}

\def\be{\begin{eqnarray}}
\def\ee{\end{eqnarray}}

\def\l[{\phantom.[}

\hoffset= - 1.0in         % switch off for draft style
\begin{document}

\hfill ITEP/TH-35/15

\hfill IITP/TH-15/15

\hfill INR-TH/2015-027

\bigskip

\bigskip

\centerline{\Large{\textsc{Decomposing Nekrasov Decomposition
  }}}

\bigskip

\centerline{A.~Morozov$^{a,b,c,}$\footnote{morozov@itep.ru}, Y.~Zenkevich$^{a,c,d,}$\footnote{yegor.zenkevich@gmail.com}}

\bigskip

\centerline{\it $^a$ITEP, Moscow, Russia}

\centerline{\it $^b$Institute for Information Transmission Problems,
  Moscow, Russia}

\centerline{\it $^c$National Research Nuclear University MEPhI, Moscow,
  Russia}

\centerline{\it $^d$Institute for Nuclear Research of Russian Academy of
  Sciences, Moscow, Russia}

\bigskip

\centerline{\textsc{abstract}}

\bigskip

{\footnotesize AGT relations imply that the four-point conformal block
  admits a decomposition into a sum over pairs of Young diagrams of
  essentially rational Nekrasov functions --- this is immediately seen
  when conformal block is represented in the form of a matrix model.
  However, the $q$-deformation of the same block has a deeper
  decomposition --- into a sum over a quadruple of Young diagrams of a
  product of four topological vertices. We analyze the interplay
  between these two decompositions, their properties and their
  generalization to multi-point conformal blocks.  In the latter case
  we explain how Dotsenko-Fateev all-with-all (star) pair
  ``interaction'' is reduced to the quiver model nearest-neighbor
  (chain) one. We give new identities for $q$-Selberg averages of
  pairs of generalized Macdonald polynomials. We also translate the
  slicing invariance of refined topological strings into the language
  of conformal blocks and interpret it as abelianization of
  generalized Macdonald polynomials.}

\bigskip

\bigskip

\tableofcontents

\section{Introduction}

Conformal blocks \cite{CFT} are among the most interesting and
important quantities under study in modern theoretical physics.
Perturbatively they are defined as series of matrix elements in
highest weight representations of Virasoro algebra, see \cite{CBser}
for recent reviews.  Non-perturbatively they are examples of
matrix-model $\tau$-functions \cite{mamotau}, associated with peculiar
conformal \cite{confmamo} (also known as Dotsenko-Fateev \cite{DF} or
Penner \cite{Pen}) matrix models, and exhibit non-trivial and almost
unexplored behavior in various regions of moduli space \cite{ItoCB}.
Their modular transformations \cite{Modtr} are important for the study
of knot polynomials (Wilson loop averages in Chern-Simons theory
\cite{CSWit}), see \cite{GMMMS} for a recent outline.  AGT relations
\cite{AGT} connect conformal blocks to LMNS quantization \cite{LMNS}
of the Seiberg--Witten theory \cite{SW} and express them in terms of
Nekrasov functions \cite{Nek}.  Both the matrix model and Nekrasov
function formalisms imply natural lifting of original conformal blocks
to $(q,t)$-dependent quantities --- looking from different
perspectives this can be either a $\beta$- or a $q$-deformation,
associated with $5d$ generalization of Seiberg-Witten theory
\cite{5dSW}.  It is at this level that the full duality pattern gets
clear and manifest.

Finally, as a quintessence of all this, conformal blocks are
expressible through topological vertices \cite{tove} --- and this will
be the story we concentrate on in the present paper.  This relation
involves not only the full-scale theory of Schur and Macdonald
functions \cite{MD}, but also conceptually important notions of
star-chain duality and Selberg factorization. The idenitification
between $q$-deformed CFT blocks and topological vertices has been used
in~\cite{genMD} to prove the spectral duality~\cite{spec-dual} of the
former. In the present paper we generalize this identification to the
higher-point case. We also clarify the relation between preferred
direction in refined topological strings and the basis of states in
conformal field theory Hilbert space.

\subsection{Conformal blocks and characters}

\begin{figure}[h]
  \centering
  \includegraphics[width=8cm]{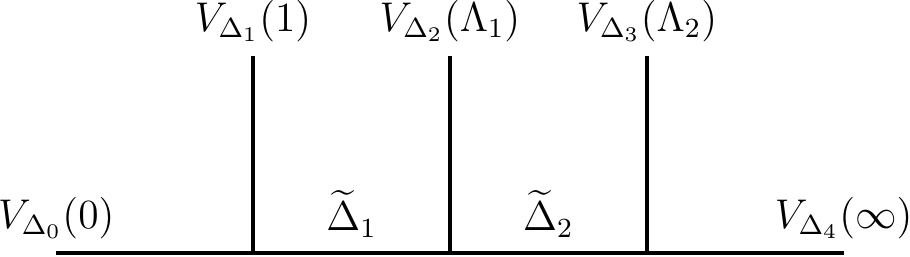}
  \caption{Comb-like 5-point conformal block on a sphere.}
  \label{fig:8}
\end{figure}

Conformal blocks are best described by the version of Dotsenko-Fateev
(DF) conformal matrix model, introduced and investigated in
\cite{MMSh}
\begin{multline}
  \mathcal{B}_{k+2}\cdot \mathcal{B}_{U(1)}= \int_0^1 d^{N_1}z
  \int_0^{\Lambda_2^{-1} }d^{N_2}z \cdots \int_0^{\Lambda_2^{-1} \
    \ldots \Lambda_{k}^{-1} }\!\! d^{N_{k}}z\ \
  \mu_{\mathrm{DF}} (z) =\\
  = \int_0^1 d^{N_1}z \int_0^{\Lambda_2^{-1} }d^{N_2}z \cdots
  \int_0^{\Lambda_2^{-1} \ \ldots \Lambda_{k}^{-1} }\!\! d^{N_{k}}z\ \
  \prod_{i\neq j} \left( 1 - \frac{z_i}{z_j} \right)^{\beta}
  \times\\
  \times \prod_{i=1}^{N_1 + \ldots + N_k} z_i^{\alpha_0} \left( 1 -
    z_i \right)^{v_1} \left( 1 - \Lambda_2 z_i \right)^{v_2} \ldots
  \left( 1 - \Lambda_2 \cdots \Lambda_k z_i
  \right)^{v_k}, \label{eq:33}
\end{multline}
where $\mathcal{B}_{U(1)}$ is an explicit function representing the
contribution of an extra free boson. We find it most convenient to use
the number $k$ of independent integration contours as a parameter ---
then what we get is a $(k+2)$-point conformal block, while the number
of bifundamentals in the gauge theory description below will be
$k-2$. The parameters of conformal block can be conveniently
summarized in a diagram, such as one shown in Fig.~\ref{fig:8}.

External dimensions
\begin{equation}
  \beta \Delta_i = v_i
  \left(v_i + \beta - 1 \right)
\end{equation}
are parameterized by the ``momenta'' $v_i$, while internal
dimensions
\begin{equation}
  \beta \widetilde{\Delta}_a = \Big(\alpha_0+ v_1+ \ldots +
  v_{a-1} + \beta N_1 + \ldots + \beta N_{a-1}\Big)
  \Big(\alpha_0+ v_1 + \ldots + v_{a-1} + \beta N_1 + \ldots + \beta
  N_{a-1} + \beta - 1\Big)
\end{equation}
are expressed through the numbers $N_a$ of screening integrations,
i.e. conformal block is considered as analytical continuation of the
integral in the number of integrations.  It is important for this
description that the integral is of Selberg type \cite{Selb} and
analytical continuation in $N_a$ is actually under control.

The next important fact \cite{ItoMAMOschur} is that the
inter-screening coupling is reduced to a square of
\begin{equation}
  \prod_{a,b} \left( 1 - \frac{x_a}{y_b} \right)^{\beta}
  = \exp \left\{-\beta \sum_k \frac{1}{k}\sum_{a,b} \left(\frac{x_a}{y_b}\right)^k\right\}
  = \exp \left\{-\beta \sum_k\frac{p_k[x]p_k[y^{-1}]}{k}\right\}
  = \sum_A  J^A[x]\, J_A^*[y]
\label{VDchar}
\end{equation}
with
\begin{equation}
  J^A[x] = J_A(p[x]) \ \ \ \ \ \ \ \ \  {\rm and} \ \ \ \ \ \ \ \ \
  J_A^*[y]= J_A(-p[y^{-1}]),
\end{equation}
i.e.\ to a bilinear combination of Jack characters $J_A$, which for
$\beta=1$ are just ordinary Schur functions
$\chi_A$. Since~(\ref{VDchar}) still needs to be squared, this reduces
the four-point conformal block to a bilinear combination of bi-character
Selberg averages~\cite{Selb} over $x$ and $y$,
\begin{equation}
\mathcal{B}_{4} = \sum_{A,B} \underbrace{\left\langle \chi^A[x]\chi^B[x]\right\rangle
\left\langle \chi_A^{*\vphantom{A}}[y]\chi_B^*[y]\right\rangle
}_{Z_{AB}}
\label{HuStr}
\end{equation}
which are exactly calculable rational combinations of $v$-parameters,
and are basically nothing but Nekrasov functions \cite{Nek}, labeled
by arbitrary pairs $A,B$ of Young diagrams.

This line of reasoning reduces AGT relation~\cite{AGT} between
conformal block and Nekrasov functions to Hubbard-Stratanovich
resummation of Selberg integrals~\cite{MMSh}.  There are important
details, making the story a little more technically involved,
especially for $\beta\neq 1$ (i.e.\ for the central charge $c \neq
1$)~\cite{betadefo},~\cite{genMD}, but in what follows we try to
separate concepts from technicalities, putting simplified general
considerations before exact, but overloaded, formulas.

\bigskip

After $q$-deformation (which in the Seiberg-Witten theory framework
means going from $4d$ to $5d$ Yang-Mills theories~\cite{5dSW}), the
integral remains basically the same, only the integration is replaced
by Jackson $q$-integration\footnote{Jackson $q$-integral is defined as
  a sum $\int_0^a f(x) d_q x = (1 - q) \sum_{k \geq 0}^\infty q^k a
  f(q^k a)$.} \cite{DF5d}:
\begin{multline} {\cal B}_{k+2}\cdot \mathcal{B}_{U(1)} = \int_0^1 d^{N_1}_q z
  \int_0^{\Lambda_2^{-1} }d^{N_2}_q z \cdots \int_0^{\Lambda_2^{-1} \
    \ldots \Lambda_{k}^{-1} }\!\! d^{N_k}_q z\ \ \prod_{i\neq
    j} \prod_{m=0}^{\beta-1} \left( 1 - q^m \frac{z_i}{z_j} \right)\times \\
  \times \prod_{i=1}^{N_1 + \ldots + N_k} \left\{z_i^{\alpha_0}
    \prod_{m=0}^{v_1 - 1} \left( 1 - q^m z_i \right) \prod_{m=0}^{v_2
      - 1} \left( 1 - q^m \Lambda_2 z_i \right)\ \ldots
    \prod_{m=0}^{v_k - 1} \left( 1 - q^m \Lambda_2 \cdots \Lambda_k
      z_i \right) \right\}
\label{CBq}
\end{multline}
Most importantly, now it acquires additional, refined decomposition
--- which for $k+2=4$ is not {\it bi}\,linear, but rather {\it
  quadri}\,linear:
\begin{equation}
  \left\langle \chi_A \chi_B \right\rangle = \sum_R S_{AR}S_{RB} \ \ \Longrightarrow \ \  \mathcal{B}_{4} \sim \sum_{Y_1, Y_2, Y_3, Y_4}
  S_{A R_1} S_{R_1 B} S_{B R_2} S_{R_2 A}
\end{equation}
--- and this is the decomposition which is related to {\it topological
  vertex}~\cite{tove},~\cite{ref-tove},~\cite{ref-tove-ak} and
geometric engineering~\cite{geing}.  The origin of two extra Young
diagrams is simple: summation over them substitutes integration over
$x$ and $y$ variables in the definition of averages in~(\ref{HuStr})
-- this appears to be the right way to interpret the multiple Jackson
integrals/sums in~(\ref{CBq}).

\subsection{Seiberg-Witten theory and topological string pattern}

\begin{figure}[h]
\centering
\includegraphics[width=15cm]{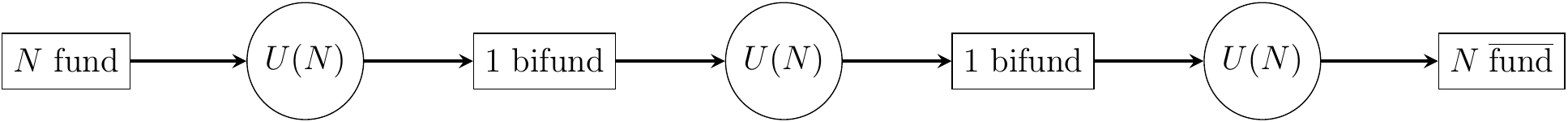}
\caption{The quiver diagram of $U(2)^3$ gauge theory.}
\label{fig:2}
\end{figure}

To better understand the origin of the multi-character decomposition
let us investigate the structure on the gauge theory side of the AGT
duality. Conformal blocks correspond to instanton partition functions
of quiver gauge theories which are given by Nekrasov formulas. The
comb-like $(k+2)$-point conformal blocks on a sphere correspond to
linear quiver theories, in which the gauge group is a product of
$(k-1)$ $U(N)$ factors and the matter content is encoded in the quiver
diagram as, e.g.\ in Fig.~\ref{fig:2}.

Here a circle is a gauge group, a box denotes a collection of matter
hypermultiplets, a outgoing (resp.\ incoming) link connecting a circle
with a box indicates that the corresponding hypermultiplets transform
as a fundamental (resp.\ antifundamental) under the gauge group. The
structure of the corresponding Nekrasov function is modelled after the
quiver diagram above:
\begin{equation}
  \label{eq:32}
  Z_{\mathrm{Nek}} = \sum_{\vec{Y}_a} \Lambda_1^{|\vec{Y}_1|} \cdots
  \Lambda_k^{|\vec{Y}_k|} z_{\mathrm{fund}}(\vec{Y}_1)
  \frac{1}{z_{\mathrm{vec}}(\vec{Y}_1)}
  z_{\mathrm{bifund}}(\vec{Y}_1, \vec{Y}_2) \cdots z_{\mathrm{bifund}}(\vec{Y}_{k-1}, \vec{Y}_k)
  \frac{1}{z_{\mathrm{vec}}(\vec{Y}_k)} z_{\overline{\mathrm{fund}}}(\vec{Y}_k),   
\end{equation}
where the definitions of the rational factors $z_{\mathrm{fund},
  \mathrm{vect},\ldots}$ are given in
Appendix~\ref{sec:five-dimens-nekr}. The structure of each term in the
decomposition is linear, in particular for a $(k+2)$-point conformal
block there are $k-1$ vector multiplet contributions and $k-2$
bifundamental matter hypermultiplets. Such quiver or chain
decomposition of the conformal block is obtained by inserting a
special basis of states $|\Delta, \vec{Y} \rangle$ labelled by a pair
of Young diagrams in the intermediate channels of the block:
\begin{equation}
  \label{eq:29}
  \mathcal{B}_{k+2} \cdot \mathcal{B}_{U(1)} = \sum_{\vec{Y}_a} \langle V_{\Delta_0}(0) V_{\Delta_1}(1)|
  \widetilde{\Delta}_1, \vec{Y}_1 \rangle \langle \widetilde{\Delta}_1, \vec{Y}_1 |
  V_{\Delta_2} (\Lambda_1)|  \widetilde{\Delta}_2, \vec{Y}_2 \rangle \cdots \langle \widetilde{\Delta}_{k-1}, \vec{Y}_{k-1} |
  V_{\Delta_k}(\Lambda_k) V_{\Delta_{k+1}} (\infty) \rangle.
\end{equation}
In the language of DF integrals this corresponds to the decomposition
of the measure $\mu_{\mathrm{DF}}(x)$ in sets of orthogonal
polynomials as in Eq.~\eqref{VDchar}. Each matrix element in
Eq.~\eqref{eq:29} is then given by the Selberg average of a collection
of orthogonal polynomials as in Eq.~\eqref{HuStr}. For $c=1$ the
special basis which reproduces the corresponding factor in the
Nekrasov function~\eqref{eq:32} is given by Schur polynomials. We will
compute the most general matrix element using $q$-Selberg averages and
show that it is indeed given by the Nekrasov expression.

For $5d$ gauge theories compactified on a circle of radius $R_{5}$ the
structure of Nekrasov function remains basically the same. The only
change is that all the monomial factors in the rational functions
$z_{\mathrm{fund}, \mathrm{vec}, ...}$ are transformed into
$q$-analogues roughly as $x \to q^x - 1$, where $q = e^{-\epsilon_2
  R_5}$. However, quite remarkably in this case Nekrasov partition
function --- or conformal block --- turns out to have yet another
interpretation. Gauge theory in five dimensions can be obtained by
compactification of M-theory on a toric Calabi--Yau
threefold. Partition function of the resulting theory is equal to the
(refined) topological string partition function, which can be computed
by the topological vertex technique as follows.

One first draws the toric diagram of the CY threefold and assigns to
each internal edge the complexified K\"ahler parameter $Q$ of the
corresponding two-cycle. One also assigns a Young diagram to each
internal edge, and an empty diagram to each external edge. There are
in general only trivalent vertices in the diagram, and to each of them
one assigns a certain function $C_{Y_1 Y_2 Y_3}(q)$ --- the
\emph{topological vertex}~\cite{tove} --- depending in a cyclically
symmetric way on three Young diagrams $Y_a$ residing on the adjacent
edges and also on the parameter $q = e^{-\epsilon_1 R_5}$:
\begin{equation}
  \label{eq:36}
   C_{ABC}(q) = \quad \parbox{2cm}{
     \includegraphics[width=2cm]{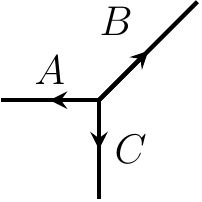}}\quad = q^{\frac{\kappa(A)}{2}} \chi_C \left(q^{\rho}\right) \sum_D  \chi_{A^{\mathrm{T}}/D} \left(q^{C+\rho}\right) \chi_{B/D}\left(q^{C^{\mathrm{T}}+\rho} \right),
\end{equation}
where $q^{\rho_i} = q^{\frac{1}{2} - i}$ and $\kappa(A) =
\sum_{(i,j)\in A} 2(j-i)$. The partition function is computed by
summing up over all the Young diagrams with weights given by the
product of all topological vertices and the ``propagators'' of the
form $(-Q)^{|Y|} f_Y(q)^n$ where $n$ is the framing factor depending
on the relative orientation of the edges adjacent to the given edge.

\begin{figure}[h]
  \centering
  \includegraphics[width=12cm]{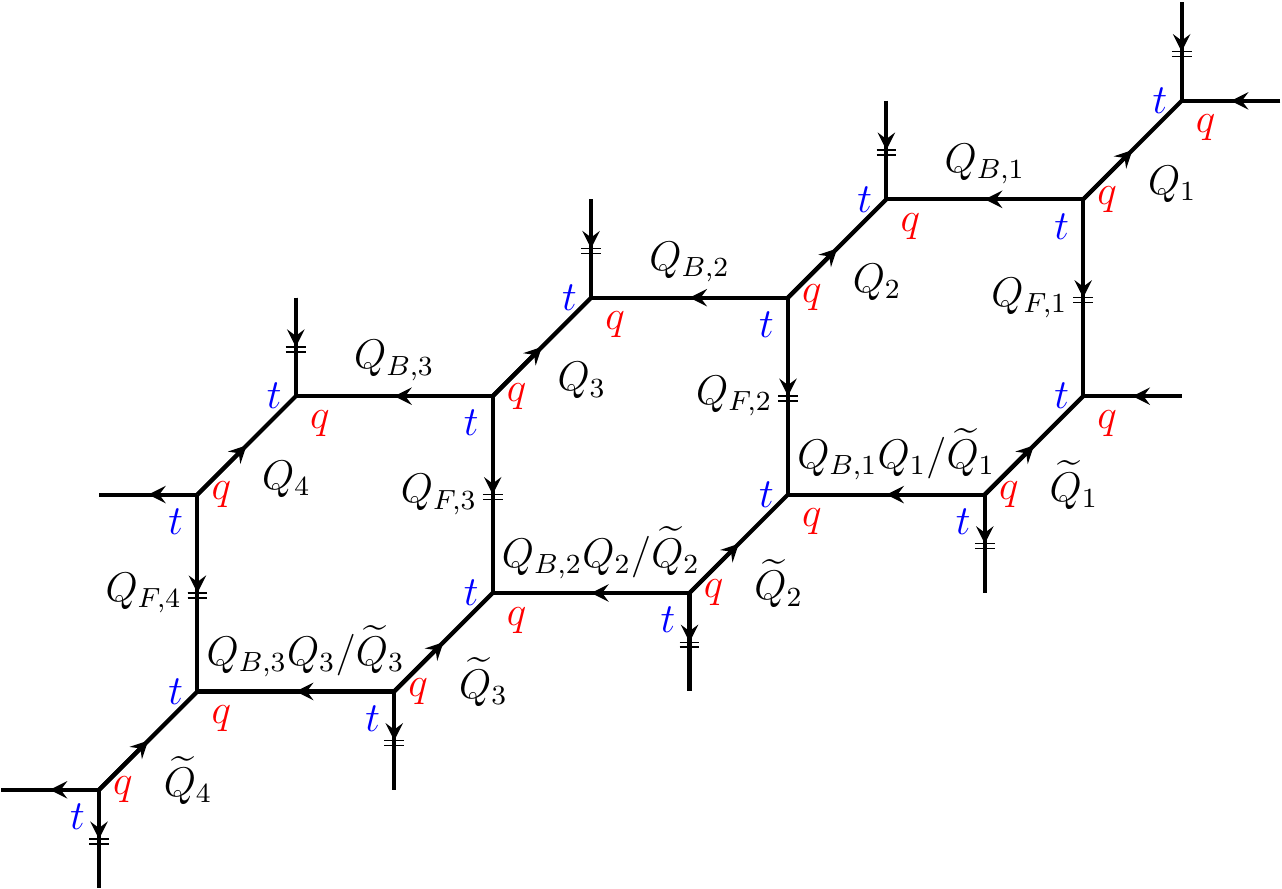}
\caption{Toric diagram corresponding to the $U(2)^3$
  gauge theory from Fig.~\ref{fig:2}.}
    \label{fig:1}
  \end{figure}

The toric diagram corresponding to a gauge theory with a product of
$k$ $U(N)$ groups is drawn using the recipe of \emph{geometric
  engineering}. It is the crossing of $N$ horizontal and $k$ vertical
lines, which intersect as shown e.g.\ in Fig.~\ref{fig:1}.

\bigskip

There is a natural decomposition of the toric diagram depicted on
Fig.~\ref{fig:1} which leads to the same quiver structure as in
Fig.~\ref{fig:2} and the Nekrasov expression~\eqref{eq:32}. One should
perform the sums over all Young diagrams except those residing on the
horizontal edges marked with $Q_{B,i}$, which are related to positions
of the vertex operators in the conformal block and the gauge theory
couplings $\Lambda_i$. In this way one obtains a sum over a chain of
\emph{pairs} of Young diagrams of certain rational factors, which turn
out to coincide with $z_{\mathrm{vect}, \mathrm{fund}, ...}$ for $t
=q$ (we introduce $t=q^{\beta}$). The resulting expression has exactly
the form of Nekrasov function~\eqref{eq:32}. Moreover, each term in
the Nekrasov decomposition can now be decomposed into an infinite sum
of simpler building blocks $Z^R_{AB}$, related to the four-point
topological string amplitude on resolved conifold. In the language of
CFT this leads to the decomposition
\begin{equation}
  \label{eq:82}
  \langle \widetilde{\Delta}_1, \vec{Y}_1 |
  V_{\Delta_2} (\Lambda_1)|  \widetilde{\Delta}_2, \vec{Y}_2 \rangle
  \sim \sum_R Z^R_{Y_{1,1} Y_{2,1}} Z^R_{Y_{1,2} Y_{2,2}},
\end{equation}
and we show that the r.h.s. is nothing but the $q$-deformed version of
DF integral expression for the matrix element in the l.h.s.

Another natural decomposition of the toric diagram --- cutting along
the \emph{vertical} edges marked with $Q_{F,i}$ (related to Coulomb
moduli of the gauge theory and intermediate dimensions in the
conformal block) --- corresponds to the \emph{spectral dual} Nekrasov
function. The gauge theory origin of this dual description is that in
$5d$ instantons are BPS particles as are the gauge bosons. Spectral
duality exchanges these two sets of BPS objects and therefore leads to
a nontrivial identification between two gauge theories. We will show
that the spectral dual decomposition of the toric diagram has a
natural interpretation in terms of DF integrals of $q$-CFT --- it is
the sum featuring in the discrete Jackson integrals, each vertical leg
corresponding to a separate integration contour
in~\eqref{eq:33}. Therefore, the spectral dual decomposition over
horizontal lines of the diagram corresponds to the DF integrals
themselves, while the original Nekrasov decomposition is the sum over
a complete set of intermediate basis states in the CFT:
\begin{equation}
  \label{eq:83}
  \mathcal{B}_{k+2} \cdot \mathcal{B}_{U(1)} \sim \sum_{R_1,\ldots,
    R_k} Z^{R_1, \ldots, R_k} Z^{R_1, \ldots, R_k}
\end{equation}

Our goal in this paper is to explain the relation between
Eq.~(\ref{CBq}), Eq.~\eqref{eq:32}, and Fig.~\ref{fig:1}. We will
learn that the identification between conformal block and Nekrasov
function requires a nontrivial rewriting of the Vandermonde
determinant (which is the product of all-with-all form) into the sum
of Nekrasov form (which is of nearest-neighbour form). We first
clarify the relation of the toric diagram and the DF integral
schematically in the simplest case of the four-point conformal block
($k=2$).  Extension to arbitrary $k$ involves an {\it a priori}
non-trivial star-chain identity, which is in fact the key to
understanding DF description of conformal blocks and relies upon the
basic properties of representation theory. Another crucial property is
Selberg factorization --- a mysterious conspiracy between the
integrands and integration measure in DF theory, between {\it what} is
averaged and {\it how} it is done. This property guarantees that the
averages of certain polynomials over the $q$-Selberg measure factorize
into products of linear factors depending on the parameters of the
integral. The last mystery is that the elementary building block in
the quadrilinear decomposition of conformal blocks, i.e.\ the
topological vertex, is closely related to the modular kernel and
therefore to certain knot polynomials.

\subsection{Refinement and slicing invariance}
\label{sec:refinement}
The calculation we have just described yields the Nekrasov function of
the $5d$ gauge theory with the particular choice of
$\Omega$-deformation parameters, i.e.\ $\epsilon_1 = - \epsilon_2$, or
equivalently $t = q$, which corresponds to $c=1$ in CFT. To obtain the
partition function in a general $\Omega$-background, one has to use
\emph{refined} topological vertex\footnote{There is a slight
  historical mismatch of notations between the refined and unrefined
  vertices. Reducing the refined vertex~\eqref{eq:75} back to the
  unrefined case to compare with Eq.~\eqref{eq:36} one needs to
  transpose all the diagrams and add some simple factors $C_{ABC}(q,q)
  = (-1)^{|A|+|B|+|C|}
  q^{\frac{\kappa(A)+\kappa(B)+\kappa(C)}{2}}C_{A^{\mathrm{T}}B^{\mathrm{T}}C^{\mathrm{T}}}(q)$.
}~\cite{ref-tove}:
\begin{multline}
  \label{eq:75}
  C_{ABC}(t,q) =\quad \parbox{2cm}{
    \includegraphics[width=2cm]{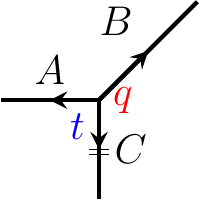}} \quad = q^{\frac{||B||^2 + ||C||^2}{2}}
  t^{-\frac{||B^{\mathrm{T}}||^2 + ||C^{\mathrm{T}}||^2}{2}} M^{(q,t)}_C
  \left(t^{-\rho}\right) \times\\
  \times\sum_D \left( \frac{q}{t}
  \right)^{\frac{|D|+|A|-|B|}{2}}  \chi_{A^{\mathrm{T}}/D}
  \left(q^{-C} t^{-\rho}\right) \chi_{B/D}\left(t^{-C^{\mathrm{T}}} q^{-\rho} \right),
\end{multline}
where $||A||^2 = \sum_i A_i^2$ and $M_C^{(q,t)}(x)$ are Macdonald
polynomials. Notice that one of the legs in the diagram is marked with
a double stroke and the other two bear $t$ and $q$ labels on
them. This is to indicate the right order of the indices and arguments
of the refined vertex, which depends on two deformation parameters and
is not cyclically symmetric as was the case for $t=q$.

The calculations generally get more technically involved, though the
strategy remains the same. The only essentially new feature in this
case is the naive loss of rotation symmetry of the diagram: the
vertical and horizontal lines are no longer equivalent. However, it
turns out that the symmetry in fact survives even for general $t$ and
$q$, though the individual vertices and propagators are not
symmetric. This statement came to be known as the slicing invariance
hypothesis. For toric geometries, which we consider, slicing
invariance is also equivalent to spectral duality~\cite{spec-dual} of
the corresponding Nekrasov partition functions, since the two sides of
the duality are related to the $\frac{\pi}{2}$ rotation of the whole
toric diagram \emph{including} the choice of preferred direction. We
look at different choices of ``slicing'' of the toric diagram and
relate them to different choices of the basis in conformal field
theory. One slicing direction corresponds to the ``naive'' basis of
Schur polynomials $\chi_A$, the other --- to the basis of generalized
Macdonald polynomials $M_{AB}$. The first set of polynomials does not
have factorized $q$-Selberg averages and does not reproduce the
Nekrasov factors, while our calculations indicate that the second one
does. Schematically
\begin{equation}
  \label{eq:80}
  \left\langle \sum_{E,F} \chi_{A/E} \chi_{C/E} \chi_{B/F} \chi_{D/F}
  \right\rangle \neq \left\langle \sum_{E,F} M_{AB/EF} M_{CD/EF}
  \right\rangle \sim \frac{z_{\mathrm{bifund}} \left([A,B], [C,D]
    \right)}{(z_{\mathrm{vect}}([A,B]) z_{\mathrm{vect}}([C,D]))^{1/2}}.
\end{equation}
The last equality is a new generalization of the ``factorization of
averages'' type of identities, studied
in~\cite{betadefo},~\cite{genMD}.

We investigate the connection between the two sets of polynomials and
introduce generalized Kostka functions $K_{AB}^{CD}$ transforming one
basis into the other:
\begin{equation}
  \label{eq:81}
  M_{AB} = \sum_{CD} K_{AB}^{CD} \chi_C \chi_D.
\end{equation}
These functions are effectively performing the $\frac{\pi}{2}$
rotation of preferred direction. In more algebraic terms they are
related to the abelianization map~\cite{Smirnov-abel} acting on the
basis in $K$-theory of instanton moduli space.  \bigskip

The paper is partitioned into a set of sections with increasing level
of detail and complexity. After reviewing the basic steps of the
construction at the simplified level in sec.~\ref{basicsteps} we fill
in the details and provide full-fledged formulas for the unrefined
case in sec.~\ref{answers}. We then treat the refined case in
sec.~\ref{sec:refin}. We provide a summary and point out future
directions in sec.~\ref{sec:discussion}.

\section{Basic steps}
\label{basicsteps}

In this section we introduce our approach to Dotsenko--Fateev integral
expansion without $q$-deformation. We consider first the most simple
example of four-point conformal block and show how decompose the
integrand in terms of Schur polynomials. Next we consider the
multi-point block and observe that a nontrivial star-chain duality is
required in this case. We demonstrate this duality explicitly using
skew Schur functions.

\subsection{Four point conformal block, no $q$-deformation}

In the case of four-point conformal block there are two contours of
integration: $\mathcal{C}_1$ stretching from $0$ to $1$ and
$\mathcal{C}_{\Lambda}$ stretching from $0$ to
$\Lambda^{-1}$. Therefore, the variables in the integration in
Eq.~\eqref{eq:33} are divided into two groups: $x_i$ and $y_i$ and the
inter-screening pairings are decomposed into a product
\begin{equation}
  \Delta(z)^{2\beta} \to \Delta(x)^{2\beta} \times \Delta(y)^{2\beta}
  \times \underbrace{\prod_{i =1}^{N_1}
    \prod_{j=1}^{N_2} \left( 1 - \frac{x_i}{y_j} \right)^{2 \beta}}_{\text{cross term}}.
  \notag
\end{equation}
The vertex operator contributions also decompose into a product of two
factors:
\begin{align}
  \prod_{i=1}^{N_1+N_2} (1 - z_i)^{v_1} &\to \prod_{i=1}^{N_1} (1 -
  x_i)^{v_1} \times \underbrace{\prod_{i=1}^{N_2} (1 -
    y_i^{-1})^{v_1}}_{\text{cross term}}, \notag\\
  \prod_{i=1}^{N_1+N_2} (1 - \Lambda z_i)^{v_2} &\to \prod_{i=1}^{N_2}
  (1 - \Lambda y_i)^{v_2} \times \underbrace{\prod_{i=1}^{N_1} (1 -
    \Lambda x_i)^{v_2}}_{\text{cross term}}. \notag
\end{align}
Making a change of variables $y_i \to \frac{y_i}{\Lambda}$ we can
write the cross terms which we denote by $\mu_{\mathrm{cross}}(x,y)$
as follows:
\begin{equation}
  \mu_{\mathrm{cross}}(x,y) = \prod_{i =1}^{N_1}
  \prod_{j=1}^{N_2} \left( 1 - \Lambda \frac{x_i}{y_j} \right)^{2
    \beta} \prod_{i=1}^{N_2} \left(1 -
    \frac{\Lambda}{y_i}\right)^{v_1} \prod_{i=1}^{N_1} (1 -
  \Lambda x_i)^{v_2} \notag
\end{equation}
Employing the Cauchy completeness identity~\eqref{VDchar} we get the
expansion of the cross contributions in terms of Jack polynomials:
\begin{multline}
  \label{eq:37}
  \mu(x,y) = \exp \left( - \sum_{n \geq 1} \frac{\Lambda^n}{n}
    ( 2 \beta p_n q_{-n} + v_1 q_{-n} + p_n v_2  )\right) =\\
  = \sum_{Y_1, Y_2} \Lambda^{|Y_1|+|Y_2|} J_{Y_1} (p_n) J_{Y_2} \left(
    -p_n - \frac{v_1}{\beta}\right) J_{Y_1} \left(-q_{-n} -
    \frac{v_2}{\beta}\right) J_{Y_2} (q_{-n}) ,
\end{multline}
where $p_n = \sum_{i=1}^{N_1} x_i^n$, $q_n = \sum_{i=1}^{N_2}
y_i^n$.

After this decomposition the DF integral becomes the double Selberg
average of Jack polynomials
\begin{equation}
  \label{eq:34}
  \mathcal{B}_4 = \sum_{Y_1, Y_2} \Lambda^{|Y_1|+|Y_2|} \left\langle J_{Y_1} (p_n) J_{Y_2} \left(-p_n -
      \frac{v_1}{\beta}\right) \right\rangle_{u_1, v_1, N_1, \beta} \left\langle J_{Y_1} \left(-q_{-n} - \frac{v_2}{\beta}\right) J_{Y_2} (q_{-n}) \right\rangle_{u_2, v_2, N_2, \beta}
\end{equation}
where the averages are taken with respect to the measure $\mu(x|u_a,
v_a, N_a, \beta) = \Delta^{2\beta}(x) \prod_{i=1}^{N_a} \left(
  x_i^{u_a} (1 - x_i)^{v_a}_{\phantom{i}}\right)$. For general $\beta$
the averages~\eqref{eq:37} do not give the Nekrasov expansion of the
conformal block (a more refined basis of \emph{generalized} Jack
functions $J_{AB}$ depending on a \emph{pair} of diagrams is
required~\cite{betadefo}). However, for the special case $\beta = 1$
when Jack polynomials turn into Schur functions the structure of
Nekrasov sum is indeed reproduced~\cite{MMSh}.

Thus, from the four-point case without $q$-deformation we learn that
decomposing the inter-screening pairings in the DF integral in terms
of characters and then taking the Selberg averages produces Nekrasov
representation of the conformal block. We now move to the multi-point
case where the star-chain duality is required to obtain Nekrasov
decomposition.

\subsection{Multi-point case. Star-chain duality}

\subsubsection{An apparent paradox}
\label{sec:an-apparent-paradox}

If one approaches the multipoint case in a naive way one arrives at
what seems to be a paradox. The DF representation contains a product
of all pairings between screening operators, i.e.\ an expression of
the form
\begin{equation}
  \label{eq:8}
  \prod_{a<b} \left(1 - \frac{x_i^a}{x_j^b}\right)^{2 \beta}.
\end{equation}
However, the gauge theory corresponding to the multipoint comb-like
conformal block is a linear quiver of the form depicted in
Fig.~\ref{fig:2}, and its Nekrasov partition function contains only
the nearest neighbour pairings:
\begin{equation}
  \label{eq:9}
  \prod_{a} z_{\mathrm{bifund}} (Y_a, Y_{a+1})
\end{equation}
Thus the multilinear decomposition of the DF integral should also have
the nearest-neighbor structure. In the four-point case there are only
two term in the product, so that all-with-all (star) type interaction
is the same as nearest-neighbour (chain) one. But how can one
decompose the multi-point product~\eqref{eq:8} into a sum of
nearest-neighbour products, how can star become equivalent to a chain?

\subsubsection{Skew characters}

The resolution of the paradox is technically based on the properties
of skew characters,
\begin{equation}
  \chi_{_{A/W}} \stackrel{\mathrm{def}}{=} \sum_B c^A_{_{BW}}\chi_{_B}
\end{equation}
where $c^A_{BC}$ are the Littlewood-Richardson coefficients,
describing multiplication of representations:
\begin{equation}
  \chi_{_B} \chi_{_C} = \sum_A c^A_{_{BC}}\chi_{_A}
  \label{multchar}
\end{equation}
\bigskip

Directly from the definitions,
\begin{equation}
  \sum_A \chi_A[x] \chi_{A/W}[y] = \sum_{A,B} c^A_{BW} \chi_A[x]\chi_B[y] =
  \chi_W[x] \cdot \sum_{B} \chi_B[x]\chi_B[y] = \chi_W[x]\cdot \prod_{i,j}(1-x_iy_j)^{-1}
\end{equation}
Moreover, this is straightforwardly generalized to
\begin{equation}
  \boxed{ \sum_A
    \chi_A\Big(p_n(x^{(1)})+\ldots+p_n(x^{(m)})\Big)\cdot \chi_{A/W}[y] =
    \chi_W\Big(p_n(x^{(1)})+\ldots+p_n(x^{(m)})\Big)\cdot \prod_{a=1}^m
    \prod_{i,j}(1-x_i^{(m)}y_j)^{-1} }
\label{sumAAW}
\end{equation}

\bigskip

Similarly,
\begin{equation}\sum_W  \chi_W[x] \chi_{A/W}[y] = \sum_{B,W} c^A_{BW} \chi_B[y]\chi_W[x]
\end{equation}
Convolution with $\chi_A[z]$ gives:
$$
\sum_A \chi_A[z] \left(\sum_W  \chi_W[x] \chi_{A/W}[y]\right)
= \sum_{B,W}\left(\sum_A c^A_{BW}\chi_A[z]\right) \chi_B[y]\chi_W[x]
= \left(\sum_B \chi_B[z]\chi_B[y]\right)\left(\sum_W\chi_W[z]\chi_W[x]\right)
$$
\vspace{-0.4cm}
\begin{equation}= \exp\left\{\sum_k \frac{1}{k}p_k(z)\Big(p_k(x)+p_k(y)\Big)\right\}
= \sum_A \chi_A[z] \chi_A\Big(p_n(x)+p_n(y)\Big)
\end{equation}
i.e.
\begin{equation}\boxed{
\sum_W  \chi_W[x] \chi_{A/W}[y] = \chi_A\Big(p(x)+p(y)\Big)
}
\label{sumWAW}
\end{equation}
At $x=y$ we can apply (\ref{multchar}) to the l.h.s. to get a doubling rule
\begin{equation}\chi_{_A}(2p_n) = \sum_B \Big(\sum_{V,W} C^A_{_{VW}}C^B_{_{VW}}\Big)\cdot\chi_{_B}(p_n)
\end{equation}
e.g.\ $\chi_{_{[1]}}(2p_n)=2\chi_{_{[1]}}(p_n),\ \ \chi_{_{[2]}}(2p_n) =
3\chi_{_{[2]}}(p_n) + \chi_{_{[11]}}(p_n),\ \ \chi_{_{[11]}}(2p_n) =
\chi_{_{[2]}}(p_n) + 3\chi_{_{[11]}}(p_n),\ \ \chi_{_{[3]}}(2p_n) =
4\chi_{_{[3]}}(p_n) + 2\chi_{_{[21]}}(p_n),\ \ \chi_{_{[21]}}(2p_n) =
2\chi_{_{[3]}}(p_n) + 6\chi_{_{[21]}}(p_n) + 2\chi_{_{[111]}}(p_n),\ \
\chi_{_{[111]}}(2p_n) = 2\chi_{_{[21]}}(p_n) + 4\chi_{_{[111]}}(p_n),\
\ldots$ which can be further promoted to tripling, quadrupling and
higher multiplication formulas.

\subsubsection{Resolution of the star/chain problem. From chain to
  star. Bifundamental kernel}
\label{sec:resolution}

We claim that the chain of skew characters indeed reproduces the
star-like structure of the DF integrand. The basic building block of
the chain decomposition is the \emph{bifundamental kernel}
\begin{equation}
  \label{eq:35}
  N_{AB}[y] = \sum_C \chi_{A/C}^{*}[y] \chi_{B/C}[y],
\end{equation}
where $\chi_Y^{*}[x] = \chi_Y[-p_n(x^{-1})]$. Two such kernels,
averaged over the Selberg measure like $\langle N_{AB}[y] N_{CD}[y]
\rangle$, correspond to a single bifundamental field in Nekrasov
partition function of the gauge theory depending on two \emph{pairs}
of diagrams $(A,B)$ and $(C,D)$. Observe that\footnote{There is
  another curious identity, which would be useful for toric blocks:
  $\sum_A Q^{|A|} N_{AA}[x] = \prod_{k \geq 1} (1 - Q^k)^{-1}
  \prod_{i,j \geq 1}(1 - Q^k x_i x_j^{-1}) $.}  $N_{\varnothing A}[x]
= \chi_A[x]$, $N_{A \varnothing}[x] = \chi_A^{*}[x]$.

We start with the case of five-point conformal block. Using the
identities from the previous section, we can rewrite the chain answer
into the Dotsenko-Fateev (star) form:
\begin{multline}
  \label{eq:11}
  \sum_{Y_1,Y_2} N_{\varnothing Y_1}[x] N_{Y_1 Y_2}[y] N_{Y_2
    \varnothing}[z] =\\
  = \sum_{Y_1,Y_2,W} \chi_{_{Y_1}} [x] \chi_{_{Y_1/W}}^{*}[y]
  \chi_{_{Y_2/W}}[y] \chi_{_{Y_2}}^{*}[z] = \prod_{i,j} \left( 1 -
    \frac{x_i}{y_j} \right) \sum_{Y_2,W} \chi_{_W} [x]
  \chi_{_{Y_2/W}}[y] \chi_{_{Y_2}}^{*} [z] = \\
  = \prod_{i,j} \left( 1 - \frac{x_i}{y_j} \right) \sum_{_{Y_2}}
  \chi_{_{Y_2}} \Big(p_n(x) + p_n(y)\Big) \chi_{_{Y_2}}^{*} [z] =
  \prod_{i,j} \left( 1 - \frac{x_i}{y_j} \right) \prod_{i,j} \left( 1
    - \frac{y_i}{z_j} \right) \prod_{i,j} \left( 1 - \frac{x_i}{z_j}
  \right)
\end{multline}
where $\chi^{*}[y] = \chi[-p_n(y^{-1})] = \chi[-p_{-n}(y)]$.

\bigskip

Similarly for a general $(k+2)$-point conformal block:
\begin{multline}
\label{eq:13}
\sum_{Y_a} N_{\varnothing Y_1} [x] \left(\prod_{a=1}^{k-2} N_{Y_a
    Y_{a+1}} [y_a]\right) N_{Y_{k-1} \varnothing}[z] = \sum_{Y_a,W_a} \chi_{_{Y_1}}[x]
\left(\prod_{a=1}^{k-2}
  \chi_{_{Y_a/W_a}}^{*}[y_a]\chi_{_{Y_{a+1}/W_a}}[y_a]\right)
\chi_{_{Y_{k-1}}}^{*}[z] = \\
=\prod_{i,j} \left( 1 - \frac{x_i}{y_{1j}} \right) \sum_{Y_a,W_a}
\chi_{_{Y_2}}\Big(p_n[x]+p_n[y_1]\Big) \left(\prod_{a=2}^{k-2}
  \chi_{_{Y_a/W_a}}^{*}[y_a]\chi_{_{Y_{a+1}/W_a}}[y_a]\right)
\chi_{_{Y_{k-1}}}^{*}[z] =
\\
\!\!\!\!\!\!\!  = \prod_{i,j} \left( 1 - \frac{x_i}{y_{1j}} \right)
\left( 1 - \frac{x_i}{y_{2j}} \right) \left( 1 - \frac{y_{1i}}{y_{2j}}
\right) \sum_{Y_a,W_a} \chi_{_{Y_3}}\Big(p_n[x]+p_n[y_1]+p_n[y_2]\Big)
\left(\prod_{a=3}^{k-2}
  \chi_{_{Y_a/W_a}}^{*}[y_a]\chi_{_{Y_{a+1}/W_a}}[y_a]\right)
\chi_{_{Y_{k-1}}}^{*}[z] = \\
= \ \ldots \ = \prod_{i,j}\left\{ \left( 1 - \frac{x_i}{z_{j}} \right)
  \prod_a \left( 1 - \frac{x_i}{y_{aj}} \right) \prod_{a<b}\left( 1 -
    \frac{y_{ai}}{y_{bj}} \right) \prod_a \left( 1 -
    \frac{y_{ai}}{z_j} \right)\right\}.
\end{multline}

\subsubsection{From star to chain}

Inverting this short derivation, we see that it is an iteration of the
two-step procedure, which starts from $m=k-1$ with
$F_{k-1}[Y]=\chi_{_{Y}}^*[z]$ and ends at $m=2$ with $x=y_0$.

In obvious notation:
\begin{multline}
\sum_{Y_m} F_m\{Y_m\}\!\cdot\!
\chi_{_{Y_{m}}}[y_0,y_1,\ldots,y_{m-1}]
= \sum_{Y_{m},W_{m-1}} \underline{ F_m\{Y_m\} \cdot \chi_{_{Y_{m}/W_{m-1}}}[y_{m-1}]} \cdot
\chi_{_{W_{m-1}}}[y_0,\ldots,y_{m-2}]
\end{multline}
Underlined piece goes directly to the chain-side of the identity,
while the remaining multi-character is combined with the next product
\begin{equation}\sum_{Z_{m-1}} \chi_{_{Z_{m-1}}}^*[y_{m-1}]\cdot \chi_{_{Z_{m-1}}}[y_0,\ldots, y_{m-2}]
\end{equation}
Since, whatever are the sets $u$ and $w$,
\begin{equation}\sum_Z \chi_{_Z}[u] \chi_{_Z}[w] \chi_{_W}[w] = \sum_{Z,Y} C_{_{ZW}}^Y \chi_{_Z}[u] \chi_{_Y}[w] =
\sum_Y \chi_{_{Y/W}}[u]\chi_{_Y}[w]
\end{equation}
we get:
\begin{multline}
\sum_{Y_{m-1}} \ \underbrace{
\sum_{Y_{m},W_{m-1}}  F_m\{Y_m\} \cdot \chi_{_{Y_{m}/W_{m-1}}}[y_{m-1}]
 \ \chi^*_{_{Y_{m-1}/W_{m-1}}} [y_{m-1}] }_{F_{m-1}\{Y_{m-1}\}}\ \cdot\
  \chi_{_{Y_{m-1}}}[y_0,\ldots,y_{m-2}] = \\
  = \sum_{Y_{m-1}} F_{m-1}\{Y_{m-1}\}\cdot
\chi_{_{Y_{m-1}}}[y_0,y_1,\ldots,y_{m-2}]
\end{multline}
and we are ready for the next iteration.

At $k=3$ this can be pictorially   represented as

\begin{picture}(300,100)(-100,-70)
\put(48,0){\vector(-1,0){46}}\put(48,0){\circle*{3}}\put(2,0){\circle*{3}}
\put(21,-45){\vector(-1,2){20}}\put(1,-5){\circle*{3}}\put(21,-45){\circle*{3}}
\put(49,-5){\vector(-1,-2){20}}\put(49,-5){\circle*{3}}\put(29,-45){\circle*{3}}
\put(1.5,-2.5){\circle{20}}
\put(-15,12){\mbox{$z=y_2$}}  \put(38,8){\mbox{$x=y_0$}}  \put(8,-60){\mbox{$y=y_1$}}
\put(77,-25){\mbox{$=$}}
\put(130,-10){\vector(-1,1){15}}\put(115,5){\circle*{3}}\put(130,-10){\circle*{3}}
\put(170,-10){\vector(-1,0){40}}\put(170,-10){\circle*{3}}
\put(130,-50){\vector(0,1){40}}\put(130,-50){\circle*{3}}
\put(171,-15){\vector(-1,-1){35}}\put(171,-15){\circle*{3}}\put(136,-50){\circle*{3}}
\put(170.5,-12.5){\circle{20}}
\put(110,10){\mbox{$z$}}
\put(173,2){\mbox{$x$}}
\put(128,-60){\mbox{$y$}}
\put(125,-2){\mbox{{\footnotesize  $Y_2$}}}
\put(147,-5){\mbox{{\footnotesize $W_1$}}}
\put(102,-35){\mbox{{\footnotesize $Y_2/W_1$}}}
\put(200,-25){\mbox{$=$}}
\put(250,-10){\vector(-1,1){15}}\put(235,5){\circle*{3}}\put(250,-10){\circle*{3}}
\put(293,-10){\vector(-1,0){43}}\put(293,-10){\circle*{3}}
\put(268,-46){\vector(-1,2){18}}\put(268,-46){\circle*{3}}
\put(293,-10){\vector(-1,-2){18}}\put(293,-10){\circle*{3}}\put(275,-46){\circle*{3}}
\put(308,5){\vector(-1,-1){15}}\put(308,5){\circle*{3}}
\put(230,10){\mbox{$z$}}
\put(310,10){\mbox{$x$}}
\put(268,-60){\mbox{$y$}}
\put(245,-2){\mbox{{\footnotesize  $Y_2$}}}
\put(290,-2){\mbox{{\footnotesize  $Y_1$}}}
\put(267,-5){\mbox{{\footnotesize $W_1$}}}
\put(230,-35){\mbox{{\footnotesize $Y_2/W_1$}}}
\put(290,-35){\mbox{{\footnotesize $Y_1/W_1$}}}
\end{picture}

\noindent
Dots here stand for characters, and arrows point from $\chi$ to
$\chi^*$.  At two steps we apply (\ref{multchar}) to substitute the
encycled product of characters by a single character.  Note that only
dots at the same place which are both either starting or end-points of
the arrows can be merged in this way.

\noindent
Likewise at $k=4$:

\begin{picture}(300,110)(20,-80)
\put(48,0){\vector(-1,0){46}}\put(48,0){\circle*{3}}\put(2,0){\circle*{3}}
\put(-6,-54){\vector(0,1){46}}\put(-6,-54){\circle*{3}}\put(-6,-8){\circle*{3}}
\put(56,-8){\vector(0,-1){46}}\put(56,-8){\circle*{3}}\put(56,-54){\circle*{3}}
\put(48,-62){\vector(-1,0){46}}\put(48,-62){\circle*{3}}\put(2,-62){\circle*{3}}
\put(52,-58){\vector(-1,1){54}}\put(52,-58){\circle*{3}}\put(-2,-4){\circle*{3}}
\put(52,-4){\vector(-1,-1){54}}\put(52,-4){\circle*{3}}\put(-2,-58){\circle*{3}}
\put(-2,-4){\circle{20}}  \put(0,-60){\circle{12}}
\put(-15,12){\mbox{$z=y_3$}}  \put(38,8){\mbox{$x=y_0$}}
\put(-15,-70){\mbox{$y_2$}} \put(55,-70){\mbox{$y_1$}}
\put(77,-25){\mbox{$=$}}
\put(130,-10){\vector(-1,1){15}}\put(115,5){\circle*{3}}\put(130,-10){\circle*{3}}
\put(156,-10){\vector(-1,0){26}}\put(156,-10){\circle*{3}}
\put(130,-52){\vector(0,1){42}}\put(130,-52){\circle*{3}}
\put(155,-17){\vector(-1,-2){20}}\put(155,-17){\circle*{3}}\put(135,-57){\circle*{3}}
\put(194,-3){\vector(0,-1){61}}\put(194,-3){\circle*{3}}\put(194,-64){\circle*{3}}
\put(186,5){\vector(-2,-1){30}}\put(186,5){\circle*{3}}
\put(190,1){\vector(-2,-1){35}}\put(190,1){\circle*{3}}
\put(186,-40){\vector(-1,1){30}}\put(186,-40){\circle*{3}}
\put(190,-52){\vector(-1,1){35}}\put(190,-52){\circle*{3}}
\put(155.5,-13.5){\circle{16}}\put(188,4){\circle{12}}\put(188,-46){\circle{18}}
\put(110,10){\mbox{$z$}}
\put(196,5){\mbox{$x$}}
\put(125,-67){\mbox{$y_2$}}
\put(199,-60){\mbox{$y_1$}}
\put(123,0){\mbox{{\footnotesize  $Y_3$}}}
\put(137,-5){\mbox{{\footnotesize $W_2$}}}
\put(102,-35){\mbox{{\footnotesize $Y_3/W_2$}}}
\put(214,-25){\mbox{$=$}}
\put(260,-10){\vector(-1,1){15}}\put(245,5){\circle*{3}}\put(260,-10){\circle*{3}}
\put(290,-10){\vector(-1,0){30}}\put(290,-10){\circle*{3}}
\put(260,-56){\vector(0,1){46}}\put(260,-56){\circle*{3}}
\put(290,-10){\vector(-1,-2){23}}\put(267,-56){\circle*{3}}
\put(310,-10){\vector(-1,0){20}}\put(310,-10){\circle*{3}}
\put(333,-56){\vector(-1,2){23}}\put(333,-56){\circle*{3}}
\put(344,-1){\vector(0,-1){55}}\put(344,-1){\circle*{3}}\put(344,-56){\circle*{3}}
\put(340,5){\vector(-2,-1){30}}\put(340,5){\circle*{3}}
\put(342,2){\circle{15}}
\put(240,10){\mbox{$z$}}
\put(350,5){\mbox{$x$}}
\put(260,-67){\mbox{$y_2$}}
\put(335,-67){\mbox{$y_1$}}
\put(253,0){\mbox{{\footnotesize  $Y_3$}}}
\put(270,-5){\mbox{{\footnotesize $W_2$}}}
\put(295,-5){\mbox{{\footnotesize $Y_2$}}}
\put(315,0){\mbox{{\footnotesize  $W_1$}}}
\put(232,-40){\mbox{{\footnotesize $Y_3/W_2$}}}
\put(275,-50){\mbox{{\footnotesize $Y_2/W_2$}}}
\put(292,-33){\mbox{{\footnotesize $Y_2/W_1$}}}
\put(365,-25){\mbox{$=$}}
\put(410,-10){\vector(-1,1){15}}\put(395,5){\circle*{3}}\put(410,-10){\circle*{3}}
\put(440,-10){\vector(-1,0){30}}\put(440,-10){\circle*{3}}
\put(410,-56){\vector(0,1){46}}\put(410,-56){\circle*{3}}
\put(440,-10){\vector(-1,-2){23}}\put(417,-56){\circle*{3}}
\put(460,-10){\vector(-1,0){20}}\put(460,-10){\circle*{3}}
\put(483,-56){\vector(-1,2){23}}\put(483,-56){\circle*{3}}
\put(490,-10){\vector(-1,0){30}}\put(490,-10){\circle*{3}}
\put(490,-10){\vector(0,-1){46}}\put(490,-56){\circle*{3}}
\put(505,5){\vector(-1,-1){15}}\put(505,5){\circle*{3}}
%\put(492,2){\circle{15}}
\put(390,10){\mbox{$z$}}
\put(505,10){\mbox{$x$}}
\put(410,-67){\mbox{$y_2$}}
\put(483,-67){\mbox{$y_1$}}
\put(403,0){\mbox{{\footnotesize  $Y_3$}}}
\put(420,-5){\mbox{{\footnotesize $W_2$}}}
\put(447,-5){\mbox{{\footnotesize $Y_2$}}}
\put(470,-5){\mbox{{\footnotesize  $W_1$}}}
\put(488,0){\mbox{{\footnotesize  $Y_1$}}}
\put(382,-40){\mbox{{\footnotesize $Y_3/W_2$}}}
\put(425,-50){\mbox{{\footnotesize $Y_2/W_2$}}}
\put(442,-33){\mbox{{\footnotesize $Y_2/W_1$}}}
\put(493,-40){\mbox{{\footnotesize $Y_1/W_1$}}}
\end{picture}

\noindent
and at $k=5$:

\begin{picture}(300,300)(-100,-270)
\put(252,-32){\circle{14}}
  \put(-12,0){\vector(-2,-1){40}}\put(-12,0){\circle*{3}}\put(-52,-20){\circle*{3}}
\put(-4,0){\vector(-1,-4){27}}\put(-4,0){\circle*{3}}\put(-31,-108){\circle*{3}}
\put(4,0){\vector(1,-4){27}}\put(4,0){\circle*{3}}\put(31,-108){\circle*{3}}
\put(12,0){\vector(2,-1){40}}\put(12,0){\circle*{3}}\put(52,-20){\circle*{3}}
\put(52,-28){\vector(-1,0){104}}\put(52,-28){\circle*{3}}\put(-52,-28){\circle*{3}}
\put(52,-36){\vector(-1,-1){76}}\put(52,-36){\circle*{3}}\put(-24,-112){\circle*{3}}
\put(54,-44){\vector(-1,-4){16}}\put(54,-44){\circle*{3}}\put(38,-108){\circle*{3}}
\put(24,-112){\vector(-1,1){76}}\put(24,-112){\circle*{3}}\put(-52,-36){\circle*{3}}
\put(19,-117){\vector(-1,0){38}}\put(19,-117){\circle*{3}}\put(-19,-117){\circle*{3}}
\put(-38,-108){\vector(-1,4){16}}\put(-38,-108){\circle*{3}}\put(-54,-44){\circle*{3}}
\put(-12,8){\mbox{$y_0=x$}}
\put(-90,-15){\mbox{$y_4=z$}}
\put(60,-32){\mbox{$y_1$}}
\put(-42,-123){\mbox{$y_3$}}
\put(28,-120){\mbox{$y_2$}}
\put(-52,-32){\circle{34}}
\put(-23,-111){\circle{22}}
\put(90,-70){\mbox{$=$}}
\put(180,12){\vector(-1,-1){40}}\put(180,12){\circle*{3}}\put(-52,-20){\circle*{3}}
%\put(196,0){\vector(-1,-4){27}}\put(-4,0){\circle*{3}}\put(-31,-108){\circle*{3}}
\put(183,8){\vector(-1,-1){44}}\put(183,8){\circle*{3}}\put(-52,-20){\circle*{3}}
\put(203,4){\vector(1,-4){28}}\put(203,4){\circle*{3}}\put(231,-108){\circle*{3}}
\put(212,0){\vector(2,-1){40}}\put(212,0){\circle*{3}}\put(252,-20){\circle*{3}}
\put(252,-28){\vector(-1,0){110}}\put(252,-28){\circle*{3}}\put(140,-28){\circle*{3}}
\put(252,-36){\vector(-1,0){112}}\put(252,-36){\circle*{3}}\put(140,-36){\circle*{3}}
\put(254,-44){\vector(-1,-4){16}}\put(254,-44){\circle*{3}}\put(238,-108){\circle*{3}}
\put(140,-28){\vector(-1,0){16}}
\put(224,-112){\vector(-1,1){84}}\put(224,-112){\circle*{3}}\put(-52,-36){\circle*{3}}
\put(219,-116){\vector(-1,1){80}}\put(219,-116){\circle*{3}}\put(-19,-117){\circle*{3}}
\put(164,-108){\vector(-1,2){40}}\put(164,-108){\circle*{3}}\put(124,-28){\circle*{3}}
\put(140,-36){\vector(1,-2){38}} \put(178,-112){\circle*{3}}
%
%\put(139,-36){\vector(1,-2){37}}
\put(124,-28){\vector(-1,1){20}}\put(104,-8){\circle*{3}}
%\put(169,-36){\vector(0,-1){72}}
%
\put(193,8){\mbox{$x$}}
\put(100,-4){\mbox{$z$}}
\put(260,-32){\mbox{$y_1$}}
\put(164,-123){\mbox{$y_3$}}
\put(228,-120){\mbox{$y_2$}}
\put(140,-32){\circle{20}}
\put(221.5,-114){\circle{12}}
\put(181.5,10){\circle{12}}
\put(290,-70){\mbox{$=$}}
\put(-30,-200){\vector(-1,1){20}}\put(-30,-200){\circle*{3}}\put(-50,-180){\circle*{3}}
\put(0,-200){\vector(-1,0){30}}\put(0,-200){\circle*{3}}
\put(20,-200){\vector(-1,0){20}}\put(20,-200){\circle*{3}}
\put(50,-200){\vector(-1,0){30}}\put(50,-200){\circle*{3}}
\put(70,-180){\vector(-1,-1){20}}\put(70,-180){\circle*{3}}
\put(80,-180){\vector(-1,-1){28}}\put(80,-180){\circle*{3}}
\put(90,-180){\vector(1,-1){16}}\put(90,-180){\circle*{3}}\put(106,-196){\circle*{3}}
\put(100,-200){\vector(-1,0){50}}\put(100,-200){\circle*{3}}
\put(104,-208){\vector(-1,0){52}}\put(104,-208){\circle*{3}}\put(52,-208){\circle*{3}}
\put(-19,-244){\vector(-1,4){11}}\put(-19,-244){\circle*{3}}
\put(0,-200){\vector(-1,-4){11}}\put(-11,-244){\circle*{3}}
\put(31,-244){\vector(-1,4){11}}\put(31,-244){\circle*{3}}
\put(52,-208){\vector(-1,-4){9}}\put(43,-244){\circle*{3}}
\put(51.5,-204){\circle{15}}
\put(102,-204){\circle{14}}
\put(75,-180){\circle{15}}
\put(-55,-175){\mbox{$z$}}
\put(82,-170){\mbox{$x$}}
\put(113,-202){\mbox{$y_1$}}
\put(32,-255){\mbox{$y_2$}}
\put(-18,-255){\mbox{$y_3$}}
\put(-80,-220){\mbox{$=$}}
\put(145,-220){\mbox{$=$}}
\put(190,-200){\vector(-1,1){20}}\put(190,-200){\circle*{3}}\put(170,-180){\circle*{3}}
\put(220,-200){\vector(-1,0){30}}\put(220,-200){\circle*{3}}
\put(240,-200){\vector(-1,0){20}}\put(240,-200){\circle*{3}}
\put(270,-200){\vector(-1,0){30}}\put(270,-200){\circle*{3}}
\put(290,-200){\vector(-1,0){20}}\put(290,-200){\circle*{3}}
\put(320,-200){\vector(-1,0){30}}\put(320,-200){\circle*{3}}
\put(340,-180){\vector(-1,-1){20}}\put(340,-180){\circle*{3}}
%\put(300,-180){\vector(-1,-1){28}}\put(300,-180){\circle*{3}}
%\put(310,-180){\vector(1,-1){16}}\put(310,-180){\circle*{3}}\put(326,-196){\circle*{3}}
%\put(320,-200){\vector(-1,0){50}}\put(320,-200){\circle*{3}}
%\put(324,-208){\vector(-1,0){52}}\put(324,-208){\circle*{3}}\put(272,-208){\circle*{3}}
%
\put(201,-244){\vector(-1,4){11}}\put(201,-244){\circle*{3}}
\put(220,-200){\vector(-1,-4){11}}\put(209,-244){\circle*{3}}
\put(251,-244){\vector(-1,4){11}}\put(251,-244){\circle*{3}}
\put(270,-200){\vector(-1,-4){11}}\put(259,-244){\circle*{3}}
\put(301,-244){\vector(-1,4){11}}\put(301,-244){\circle*{3}}
\put(320,-200){\vector(-1,-4){11}}\put(309,-244){\circle*{3}}
\put(165,-175){\mbox{$z$}}
\put(342,-175){\mbox{$x$}}
\put(302,-255){\mbox{$y_1$}}
\put(252,-255){\mbox{$y_2$}}
\put(202,-255){\mbox{$y_3$}}
\put(178,-187){\mbox{{\footnotesize $Y_4$}}}
\put(201,-195){\mbox{{\footnotesize $W_3$}}}
\put(227,-195){\mbox{{\footnotesize $Y_3$}}}
\put(251,-195){\mbox{{\footnotesize $W_2$}}}
\put(277,-195){\mbox{{\footnotesize $Y_2$}}}
\put(301,-195){\mbox{{\footnotesize $W_1$}}}
%\put(203,-195){\mbox{{\footnotesize $Y$}}}
\put(322,-187){\mbox{{\footnotesize $Y_1$}}}
\put(170,-230){\mbox{{\footnotesize $Y_4/W_3$}}}
\put(213,-237){\mbox{{\footnotesize $Y_3/W_3$}}}
\put(220,-225){\mbox{{\footnotesize $Y_3/W_2$}}}
\put(263,-237){\mbox{{\footnotesize $Y_2/W_2$}}}
\put(270,-225){\mbox{{\footnotesize $Y_2/W_1$}}}
\put(313,-237){\mbox{{\footnotesize $Y_1/W_1$}}}
\end{picture}

The main secret behind this derivation is that the structure constants
in (\ref{multchar}) are always the same --- do not depend on the
number of ``Miwa variables'' $y_i$ in $\chi[y_0,\ldots,y_m]$ --- what
allows to merge entire collections of points and parallel arrows in
above examples. This conspiracy between characters and the structure
constants adds to associativity of multiplication and together they
provide the star-chain equivalence.

\subsection{Factorization of Selberg averages}

The ``chain'' decomposition of DF integrals~\eqref{eq:13} is also tied
with the structure of the Selberg averages. More concretely, the
averages of the bifundamental kernels~\eqref{eq:35} are given by the
\emph{factorized} formulas:
\begin{equation}
  \label{eq:40}
  \left\langle N_{AC}[x] N_{BD}[-p_n(x)-v] \right\rangle_{u,v,N, \beta=1} =
  (-1)^{|B|+|D|} \frac{ z_{\mathrm{bifund}}^{4d}\left( [A,B],[C,D],
      u/2 + v/2 + N, u/2, -v/2 \right)}{C_A^{4d} C_B^{4d} C_C^{4d} C_D^{4d} G_{AB}^{4d}(u + v + 2N) G_{CD}^{4d}(u) },
\end{equation}
where
\begin{gather}
\label{eq:39}
C_A^{4d} = \prod_{(i,j) \in A} \left(\mathrm{Arm}_{A}(i,j) +
  \mathrm{Leg}_{A}(i,j) + 1\right),\\
G_{AB}^{4d}(x) = \prod_{(i, j) \in A} \left( x + A_i - j +
  B^{\mathrm{T}}_j - i + 1 \right) \prod_{(i,j) \in B}\left(x -B_i +
  j - 1 -A^{\mathrm{T}}_j + i \right).\\
z_{\mathrm{bifund}}^{4d}\left( \vec{A}, \vec{B}, \vec{a},\vec{b},
  m\right) = \prod_{i,j=1}^2 G^{4d}_{A_i B_j}(a_i - b_j - m)
\end{gather}
and $\vec{a} = (a,-a)$, $\vec{b} = (b,-b)$. This factorization means
that expansion of the DF integrand in terms of the bifundamental
kernels $N_{AB}$ indeed reproduces the Nekrasov decomposition. In the
next section we will compute the $q$-deformed averages and show how to
decompose them even further to obtain topological vertices.

\section{Complete formulas for $t=q$}
\label{answers}

In this section we flesh out the basic formulas introduced in the
previous section and incorporate $q$-deformation into our
framework. After $q$-deformation, we obtain the natural identification
of DF integral decompositions with topological vertices. We start with
especially symmetric example of the four-point conformal block of
$q$-Virasoro algebra and then consider multi-point blocks. We
calculate $q$-deformed Selberg averages of the skew characters and
identify the elements of the multilinear decomposition with
topological string amplitudes.

\subsection{Four point conformal block ($t=q$)}
\label{sec:four-point-conformal}

\begin{figure}[h]
  \centering
  \includegraphics[width=6cm]{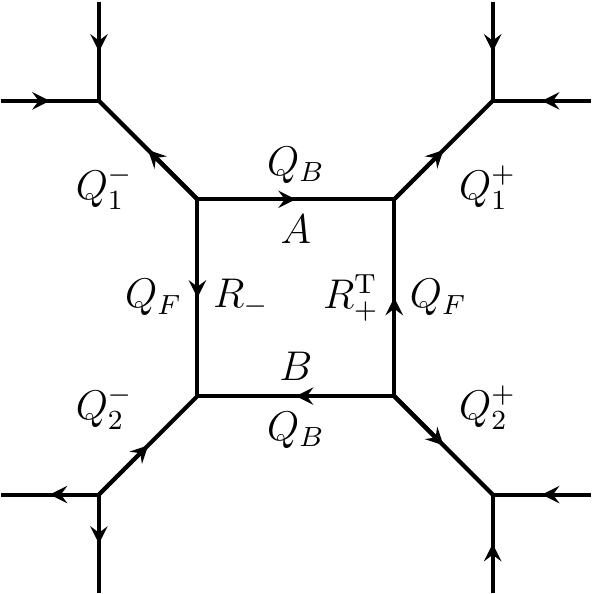}
  \caption{Toric diagram corresponding to the four-point $q$-deformed
    conformal block and $5d$ $U(2)$ gauge theory with four
    fundamentals.}
  \label{fig:3}
\end{figure}

The origin of the quadrilinear expansion of the four point conformal
block is straightforward to see: two diagrams come from the character
decomposition and two more represent the two integration contours in
the DF integral (which becomes a sum in the $q$-deformed
case~\cite{genMD}). The corresponding toric diagram is depicted in
Fig.~\ref{fig:3} and the four diagrams are denoted by $A$, $B$,
$R_{+}$ and $R_{-}$.

The DF representation is given by the sum over DF poles labelled by
two partitions $R_{\pm}$:
\begin{equation}
  \label{eq:1}
  \mathcal{B}_{4} = \sum_{R_{\pm}} \mu(x_{R_{+}})
  \mu_{\mathrm{cross}}(x_{R_{+}}, x_{R_{-}}) \mu(x_{R_{+}}).
\end{equation}
The contribution of the $q$-Selberg measure $\mu(x|u,v,N,q,q) =
\Delta(x) \prod_{i=1}^N \left( x_i^u \prod_{k=0}^{v-1} (1 - q^k x_i)
\right)$ can be evaluated explicitly and is written as follows:
\begin{gather}
  \label{eq:2}
  \frac{\mu(x_R|u,v,N,q,q)}{\mu(x_{\varnothing}|u,v,N,q,q)}= q^{|R|(2
    N + u + v - 1)} \chi_R^{(N)}(q^{\rho}) \chi_R^{(N+v)}(q^{\rho}) =
  q^{|R|(u
    - 1)} \chi_R^{(N)}(q^{-\rho}) \chi_R^{(N+v)}(q^{-\rho}) ,\\
  x_{R,i} = q^{R_i - i + N + 1},
\end{gather}
and $\chi_A^{(N)}(x)$ denotes Schur polynomial in $N$ variables
$x_i$. The cross contribution reads
\begin{multline}
  \label{eq:3}
  \mu_{\mathrm{cross}}(x,y) = \prod_{i=1}^{N_{+}} \prod_{j=1}^{N_{-}}
  \left(1 - \Lambda \frac{x_i}{y_j} \right)^2 \prod_{j=1}^{N_{-}}
  \prod_{n=0}^{v_{+}-1} \left( 1 - \Lambda q^{-n} \frac{1}{y_j}
  \right) \prod_{i=1}^{N_{+}} \prod_{l=0}^{v_{-}-1}
  \left(1 - \Lambda q^l x_i \right)=\\
  =\exp \left\{ \sum_{n \geq 1} \frac{ \Lambda^n}{n} \left[ \left( p_n
        + \frac{1 - q^{-n v_{+}}}{1 - q^{-n}} \right) (-q_{-n}) + p_n
      \left( - q_{-n} - \frac{1 - q^{n
            v_{-}}}{1 - q^n} \right)  \right] \right\} =\\
  = \sum_{A, B} \left(- \Lambda q^{N_{+} - N_{-}}\right)^{|A|+|B|}
  \chi_A \left( q^{- (N_{+} + 1/2)n} p_n +
    \sum_{i=N_{+}+1}^{N_{+}+v_{+}} q^{n(1/2-i)} \right) \chi_B
  \left( q^{- (N_{+} + 1/2)n} p_n \right)   \times\\
  \times \chi_{A^{\mathrm{T}}} \left( q^{n (N_{-} + 1/2)} q_{-n}
  \right) \chi_{B^{\mathrm{T}}} \left( q^{ (N_{-} + 1/2)n} q_{-n} +
    \sum_{i=N_{-}+1}^{N_{-}+v_{-}} q^{ - n(1/2-i)} \right),
\end{multline}
where $p_n = \sum_{i=1}^{N_{+}} x_i^n$, $q_n = \sum_{j=1}^{N_{-}}
y_j^n$ and we have employed the Cauchy identity and the identity
$\chi_A(-p_n) = (-1)^{|A|} \chi_{A^{\mathrm{T}}}(p_n) $. Collecting all the
contributions one obtains
\begin{multline}
  \label{eq:4}
  \mathcal{B}_4 = \mu(x_{\varnothing}|u_{+}, v_{+}, N_{+}) \,
  \mu(x_{\varnothing}|u_{-}, v_{-},
  N_{-})\times\\
  \times \sum_{R_{+}, R_{-}} \sum_{A,B} \left(-\Lambda q^{N_{+} -
      N_{-}}\right)^{|A|+|B|} q^{|R_{+}|(u_{+} + v_{+} + 2N_{+} - 1) +
    |R_{-}|(u_{-} + v_{-} + 2 N_{-} - 1)} \chi_{R_{+}}^{(N_{+})}
  (q^{\rho}) \chi_{R_{+}}^{(N_{+}+v_{+})} (q^{\rho})\times\\
  \times \chi_{R_{-}}^{(N_{-})} (q^{\rho}) \chi_{R_{-}}^{(N_{-}+v_{-})}
  (q^{\rho}) \chi_B^{(N_{+})} (q^{R_{+}+\rho})
  \chi_{A^{\mathrm{T}}}^{(N_{-})} (q^{-R_{-}-\rho}) \chi_A^{(N_{+}+v_{+})}
  (q^{R_{+}+\rho}) \chi_{B^{\mathrm{T}}}^{(N_{-}+v_{-})}
  (q^{-R_{-}-\rho}).
\end{multline}

Using the identities
\begin{equation}
  \label{eq:85}
  \chi_R^{(N)}(q^{\rho}) =  (-1)^{|R|} q^{-N|R|}
\chi_{R^{\mathrm{T}}}^{(-N)} (q^{\rho}),\qquad \qquad  \chi_A^{(N)}(q^{-R - \rho}) =
(-1)^{|A|} \chi_{A^{\mathrm{T}}}^{(-N)} (q^{R^{\mathrm{T}} + \rho})\notag
\end{equation}

one gets
\begin{multline}
  \label{eq:5}
  \mathcal{B}_4 = \mu(x_{\varnothing}|u_{+}, v_{+}, N_{+}) \,
  \mu(x_{\varnothing}|u_{-}, v_{-}, N_{-}) \sum_{R_{+}, R_{-}}
  \sum_{A,B} \left(\Lambda q^{N_{+} - N_{-}}\right)^{|A|+|B|}
  q^{|R_{+}|(u_{+} + v_{+} + 2N_{+} - 1) +
    |R_{-}|(u_{-} - 1)}\times \\
  \times S_{R_{+} B} (q^{-N_{+}}) S_{A R_{-}^{\mathrm{T}}} (q^{N_{-}})
  S_{R_{-}^{\mathrm{T}} B} (q^{N_{-}+v_{-}}) S_{A R_{+}} (q^{-N_{+} -
    v_{+}}),
\end{multline}
where
\begin{equation}
  \label{eq:86}
  S_{AB}(Q) = \chi_A(p_n(q^{\rho}) - p_n(Q q^{\rho}))
  \chi_B(p_n(q^{A+\rho}) - p_n(Q q^{\rho}))
\end{equation}
is the open topological string amplitude for the resolved
conifold. Pictorially $S_{AB}(Q)$ is given by one corner of the toric
diagram from Fig.~\ref{fig:3}. It is also equal to the Chern-Simons
(or WZW) $S$-matrix.

Using the AGT relations~\eqref{eq:41} one immediately obtains the
identification between the K\"ahler parameters of the toric
Calabi-Yau, CFT and the gauge theory parameters:
\begin{gather}
  \label{eq:7}
  Q_B = \Lambda q^{N_{+} - N_{-}} = \Lambda q^{-2a +m_2^{+} -
    m_2^{-}},\notag\\
  Q_F = q^{2N_{+} + u_{+} + v_{+} - 1} = q^{u_{-} - 1} = q^{-2a -1},\notag\\
  Q_1^{+} =  q^{- N_{+}
    - v_{+}} = q^{a + m_1^{+}}, \quad Q_2^{+} = q^{- N_{+}} = q^{a - m_2^{+}},\notag\\
  Q_1^{-} = q^{N_{-}} = q^{a + m_2^{-}}, \quad Q_2^{-} = q^{N_{-} +
    v_{-}} = q^{a - m_1^{-}}.\notag
\end{gather}

Let us state once more the result for the four-point $q$-deformed
conformal block for $t=q$. This block can be simultaneously decomposed
in two ways: DF integral and the decomposition in terms of the
complete basis of states. Using the simplest choice of basis states
(Schur functions) one gets a symmetric \emph{quadri}linear
decomposition in terms of characters. Moreover, this decomposition is
naturally identified with the corresponding topological string
amplitude, computed using the topological vertex technique. We now
move on to describe the multipoint case.

\subsection{Multipoint ($t=q$)}

In section~\ref{sec:resolution} we understood the star-chain
transformation for ordinary DF integrals. The $q$-deformed case goes
along the same lines. Also, as in the four-point case above, we obtain
a natural interpretation of the objects featuring in the decomposition
from the point of view of the topological strings.

\begin{figure}[h]
  \centering
  \begin{equation}
    Z \left( \left.
        \begin{smallmatrix}
          & \varnothing & \\
          Y_{a,1} & & Y_{a-1,1}\\
          Y_{a,2} & & Y_{a-1,2}\\
          & \varnothing &
        \end{smallmatrix}
      \right| Q_1, Q_2, Q_F, q, q \right) \quad = \quad \parbox{3.5cm}{\includegraphics[width=3.5cm]{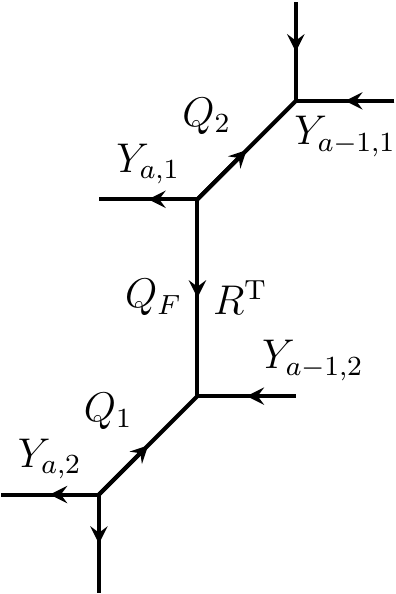}
    }\notag
  \end{equation}
  \caption{Toric diagram corresponding to bifundamental field in the
    Nekrasov function. It is also equal to $q$-deformed Selberg
    average of two bifundamental kernels as in Eq.~\eqref{eq:52}.}
  \label{fig:4}
\end{figure}

Using the star-chain relation we rewrite the Vandermonde measure in
the multipoint DF integral as a sum over chains of skew Schur
functions. We consider the corresponding expansion of the $k$-point DF
integral in terms of $k$ $q$-deformed Selberg averages of skew Schur
functions (cf.\ Eq.~\eqref{eq:13}):
\begin{equation}
  \label{eq:52}
  \mathcal{B}_{k+2} \mathcal{B}_{U(1)} = \sum_{\{\vec{Y}_a\}} \prod_{a=1}^k
  \Lambda_a^{|\vec{Y}_a|}  \Biggl\langle
    N_{Y_{a-1,1} Y_{a,1}}[x] N_{Y_{a-1,2} Y_{a,2}} \left[-p_n - \frac{1 -
        q^{-nv_a}}{1 - q^{-n}}\right] \Biggr \rangle_{u_a , v_a, N_a,q,q}
\end{equation}
where $\langle \ldots \rangle_{u_a , v_a, N_a,q,q}$ denotes the
$q$-Selberg average with the corresponding parameters and $\vec{Y}_0 =
\vec{Y}_k = (\varnothing, \varnothing)$. Each average depends on two
pairs of Young diagrams $\vec{Y}_{a-1}$, $\vec{Y}_a$ and corresponds
to the element of the toric diagram depicted on
Fig.~\ref{fig:4}. Notice that a certain $U(1)$ factor appears in the
left hand site. It can be nicely eliminated in the four-point case,
though not for higher multipoints.

In the next section we show that each average in Eq.~\eqref{eq:52}
indeed reproduces the bifundamental part of the Nekrasov function. The
whole sum thus becomes the Nekrasov function for linear quiver gauge
theory, of the form depicted in Fig.~\ref{fig:2}.

\subsection{Factorization of averages}
\label{sec:fact-aver}

Let us check that the DF averages of the bifundamental kernel
$N_{AB}[x]$ indeed factorizes. To do it we use the loop equations,
which are given in Appendix~\ref{sec:loop-equat-matr}. We obtain a
formula for the average of four Schur polynomials:
\begin{multline}
  \label{eq:15}
  \left\langle N_{AC}[x] N_{BD}\left[-p_n(x) - \frac{1 - q^{-nv}}{1 -
        q^{-n}}\right] \right\rangle_{u,v,N, q,q} =\\
  = q^{(1-v)(|C|-|A|+|D|-|B|)} \left\langle N_{AC}[x^{-1}]
    N_{BD}\left[-p_n(x^{-1}) - \frac{1 - q^{nv}}{1 -
        q^n}\right] \right\rangle_{-u-v-2N,v,N,q,q} =\\
  = (-1)^{|B|+|D|} q^{ -u |B| - (2N+u)|C| +v |D|} q^{\kappa_{ABCD}}
  \frac{z_{\mathrm{bifund}}^{(q,q)} \left( [A,B],[C,D],
      q^{-u/2},q^{-u/2-v/2-N},q^{v/2} \right)}{C'_A(q,q) C'_B(q,q)
    C'_C(q,q) C'_D(q,q) G_{AB}^{(q,q)}(q^{-u})
    G_{CD}^{(q,q)}(q^{-u-v-2N}) },
\end{multline}
where
\begin{gather}
  \label{eq:16}
  C'_A (q,t) = \prod_{(i,j) \in A} \left(1 - q^{\mathrm{Arm}_{A}(i,j)} t^{\mathrm{Leg}_{A}(i,j) + 1}\right),\\
  \kappa_{ABCD} = - \kappa_B/2 + \kappa_C/2 + \sum_{(i,j) \in A} (i-1)
  + \sum_{(i,j) \in B} (i-1) +
  \sum_{(i,j) \in C} j + \sum_{(i,j) \in D} j \\
  \kappa_A = 2 \sum_{(i,j)\in A} (j-i)
\end{gather}
and the definition of $G_{AB}^{(q,t)}$ and
$z_{\mathrm{bifund}}^{(q,t)}$ are collected in
Appendix~\ref{sec:five-dimens-nekr}. This indeed proves that the
average of four Schur polynomials gives the bifundamental Nekrasov
contribution.

\subsection{Identification with topological strings}
\label{sec:ident-with-topol}

We would like to further decompose the $q$-Selberg average in
Eq.~\eqref{eq:15} to observe the structure of the corresponding
topological string amplitude from Fig.~\ref{fig:4}. Notice that each
average contains a product of two bifundamental kernels
$N_{AB}[x]$. This corresponds to the product of two four-point
functions each having the form:
\begin{multline}
  \label{eq:17}
  Z \left( \left.
    \begin{smallmatrix}
      & \varnothing& \\
      C & & A\\
      & R^{\mathrm{T}} &
    \end{smallmatrix}
  \right|Q_2,q,q\right) =
\quad \parbox{2.8cm}{\includegraphics[width=2.8cm]{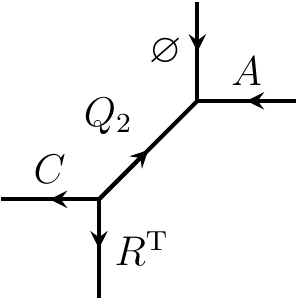}}\quad
= \sum_W (-Q_2)^{|W|}
C_{A^{\mathrm{T}}W^{\mathrm{T}}\varnothing}(q) C_{ C
W R^{\mathrm{T}}} (q) =\\
= Z \left( \left.
    \begin{smallmatrix}
      & \varnothing& \\
      \varnothing & & \varnothing\\
      & \varnothing  &
    \end{smallmatrix}
  \right|Q_2,q,q\right) q^{\frac{\kappa_C - \kappa_A}{2}} (-Q_2)^{|C|}
(-1)^{|R|} \chi_R^{(N)} (q^{-\rho})
N_{AC}^{(N)}\left[q^{-R-\rho}\right],
\end{multline}
where $Q_2 = q^N$. Recall the expression for the $q$-deformed Selberg
measure~\eqref{eq:2}, which consists of two Schur polynomials. The
product of two four-point amplitudes~\eqref{eq:17} therefore gives
exactly the product of $q$-Selberg measure with two bifundamental
kernels as in the average~\eqref{eq:52}. More explicitly, gluing two
amplitudes~\eqref{eq:17} one obtains the amplitude from
Fig.~\ref{fig:4}, which is given by:

\begin{multline}
  Z \left( \left.
        \begin{smallmatrix}
          & \varnothing & \\
          C & & A\\
          D & & B\\
          & \varnothing &
        \end{smallmatrix}
      \right| Q_1, Q_2, Q_F, q, t \right) = \sum_R (-Q_F)^{|R|} Z
    \left( \left.
    \begin{smallmatrix}
      & \varnothing & \\
      C & & A\\
      & R^{\mathrm{T}} &
    \end{smallmatrix}
\right|Q_2,q,q\right) Z \left( \left.
    \begin{smallmatrix}
      & R^{\mathrm{T}}& \\
      D & & B\\
      & \varnothing  &
    \end{smallmatrix}
  \right|Q_1,q,q\right)=\\
= Z \left( \left.
    \begin{smallmatrix}
      & \varnothing& \\
      \varnothing & & \varnothing\\
      & \varnothing  &
    \end{smallmatrix}
\right|Q_1,q,q\right) Z \left( \left.
    \begin{smallmatrix}
      & \varnothing& \\
      \varnothing & & \varnothing\\
      & \varnothing  &
    \end{smallmatrix}
  \right|Q_2,q,q\right) q^{\frac{\kappa_C - \kappa_A + \kappa_D -
    \kappa_B}{2}} (-Q_1)^{|C|} (-Q_2)^{|D|} \times\\
\times\sum_R (-Q_F)^{|R|}
s_R^{(N)}(q^{-\rho}) s_{R^{\mathrm{T}}}^{(-N-v)}(q^{-\rho}) N_{AC}^{(N)}
\left[q^{-R - \rho}\right] N_{D^{\mathrm{T}}
  B^{\mathrm{T}}}^{(-N-v)}\left[q^{-R^{\mathrm{T}}-\rho}\right] =\\
= Z \left( \left.
    \begin{smallmatrix}
      & \varnothing& \\
      \varnothing & & \varnothing\\
      & \varnothing  &
    \end{smallmatrix}
\right|Q_1,q,q\right) Z \left( \left.
    \begin{smallmatrix}
      & \varnothing& \\
      \varnothing & & \varnothing\\
      & \varnothing  &
    \end{smallmatrix}
  \right|Q_2,q,q\right) q^{\frac{\kappa_C - \kappa_A + \kappa_D -
    \kappa_B}{2}} (-Q_2)^{|C|} Q_1^{|D|} (-1)^{|B|} q^{-\left(\frac12 + N\right) (|A|-|C|+|B|-|D|)} \times\\
\times\sum_R (Q_F q^{-N-v})^{|R|} s_R^{(N)}(q^{-\rho})
s_R^{(N+v)}(q^{-\rho}) N_{AC}^{(N)} \left[q^{-R -
    \rho - \frac12 - N}\right] N_{BD}^{(N+v)}\left[-p_n \left(q^{-R-\rho- \frac12 -N}\right)\right] =\notag
\end{multline}
\begin{multline}
=Z \left( \left.
    \begin{smallmatrix}
      & \varnothing& \\
      \varnothing & & \varnothing\\
      & \varnothing  &
    \end{smallmatrix}
\right|Q_1,q,q\right) Z \left( \left.
    \begin{smallmatrix}
      & \varnothing& \\
      \varnothing & & \varnothing\\
      & \varnothing  &
    \end{smallmatrix}
  \right|Q_2,q,q\right) q^{\frac{\kappa_C - \kappa_A + \kappa_D -
    \kappa_B}{2}} (-1)^{|B|+|C|} q^{\left(2N +\frac12 \right)|C|} q^{\left(\frac12 - v\right)|D|}
q^{-\left(\frac12 +
    N\right) (|A|+|B|)} \times\\
\times  S_{u,v,N,q,q} \left\langle N_{AC}^{(N)}[x^{-1}] N_{BD}^{(N)}\left[-p_n(x^{-1}) - \frac{1
      - q^{nv}}{1 - q^n}\right] \right\rangle_{u,v,N,q,q},
\label{eq:18}
\end{multline}
where $Q_1 = q^{-N-v}$, $Q_2 = q^N$, $Q_F = q^{u+v+N - 1}$ and
$S_{u,v,N,q,q}$ is the $q$-Selberg integral without character
insertions. Thus, the DF average of four Schur functions is the same
as the topological string amplitude on two resolved conifold
geometries glued together. Moreover, we explicitly identify the sum
over intermediate states residing on the vertical edge of the diagram
on Fig.~\ref{fig:4} with the DF $q$-integration.

\begin{figure}[h]
  \centering
  \begin{equation}
    Z\left(\left.
        \begin{smallmatrix}
          &\varnothing & \varnothing & \varnothing&\\
          \varnothing & & & &\varnothing\\
          &R_3^{\mathrm{T}} & R_2^{\mathrm{T}} & R_1^{\mathrm{T}}&
        \end{smallmatrix} \right| Q_1, Q_2, Q_3, Q_{B,1}, Q_{B,2}, q,
      q \right) \quad =
    \quad \parbox{7.5cm}{\includegraphics[width=7.5cm]{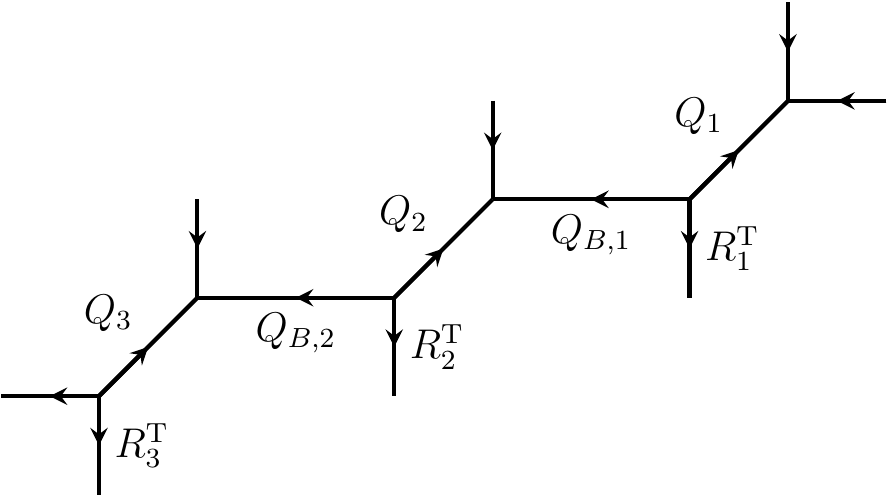}}\notag
  \end{equation}
  \caption{Toric diagram corresponding to ``half'' of the DF integrand
    for the five-point conformal block. It also give the spectral dual
    Nekrasov expansion of the corresponding linear quiver gauge theory
    with gauge group $U(2)^2$.}
  \label{fig:5}
\end{figure}

What corresponds to the whole DF \emph{integrands} on the topological
string side? One uses the star-chain relation~\eqref{eq:11} to glue
together a chain of bifundamental kernels and obtain the Vandermonde
determinant, the main constituent of the DF integrand. This
corresponds to gluing the topological string amplitudes~\eqref{eq:17}
together as depicted in Fig.~\ref{fig:5}.

Performing the sums one gets the following expression for the glued
amplitude:
\begin{multline}
  \label{eq:12}
  Z\left(\left.
      \begin{smallmatrix}
        &\varnothing & \varnothing & \varnothing&\\
        \varnothing & & & &\varnothing\\
        &R_3^{\mathrm{T}} & R_2^{\mathrm{T}} & R_1^{\mathrm{T}}&
      \end{smallmatrix} \right| Q_1, Q_2, Q_3, Q_{B,1}, Q_{B,2},
    q, q \right) =\\
  = \prod_{a=1}^3 Z \left( \left.
    \begin{smallmatrix}
      & \varnothing& \\
      \varnothing & & \varnothing\\
      & \varnothing  &
    \end{smallmatrix}
  \right|Q_a,q,q\right) \chi_{R_a}^{(N_a)} (q^{-\rho}) \sum_{A,B}
\left(Q_{B,2} q^{2N_3-N_2}\right)^{|A|}
\left(Q_{B,1}q^{2N_2-N_1}\right)^{|B|}\times\\
\times N_{\varnothing A}^{(N_3)}[x_{R_3}^{-1}]
  N_{AB}^{(N_2)}[x_{R_2}^{-1}] N_{B
    \varnothing}^{(N_1)}[x_{R_1}^{-1}] =
  q^{R_i + \rho + \frac12 + N_a} = \\
=\prod_{a=1}^3 Z \left( \left.
    \begin{smallmatrix}
      & \varnothing& \\
      \varnothing & & \varnothing\\
      & \varnothing  &
    \end{smallmatrix}
  \right|Q_a,q,q\right) \chi_{R_a}^{(N_a)} (q^{-\rho})
\prod_{i=1}^{N_1} \prod_{j=1}^{N_2} \left( 1 - \Lambda_1
  \frac{x_{R_2,j}}{x_{R_1,i}} \right) \prod_{j=1}^{N_2} \prod_{k=1}^{N_3} \left( 1 - \Lambda_2
  \frac{x_{R_3,k}}{x_{R_2,j}} \right) \prod_{i=1}^{N_1}
\prod_{k=1}^{N_3} \left( 1 - \Lambda_1 \Lambda_2
  \frac{x_{R_3,k}}{x_{R_1,i}} \right),
\end{multline}
where $x_{R_a,i} = q^{R_{a,i} + 1 + N_a - i}$, $Q_a = q^{N_a}$,
$\Lambda_1 = Q_{B,1} q^{2N_2-N_1}$ and $\Lambda_2 = Q_{B,2}
q^{2N_3-N_2}$. Eq.~\eqref{eq:12} gives exactly \emph{half} of the
terms in the DF integrand. The whole integrand consists of the
$q$-Selberg measure~\eqref{eq:2}, which includes \emph{two} Schur
functions, and the ``cross term''~\eqref{eq:3}, which is given by a
\emph{square} of the Vandermonde determinant. The other half of the
terms arises from the lower half of the toric diagram, so that the
total integrand is obtained as in Fig.~\ref{fig:1}:
\begin{multline}
  \label{eq:53}
  \mathcal{B}_5 \mathcal{B}_{U(1)} = \sum_{R_1, R_2, R_3}
  (-Q_{F,1})^{|R_1|}(-Q_{F,2})^{|R_2|}(-Q_{F,3})^{|R_3|} Z\left(\left.
      \begin{smallmatrix}
        &\varnothing & \varnothing & \varnothing&\\
        \varnothing & & & &\varnothing\\
        &R_3^{\mathrm{T}} & R_2^{\mathrm{T}} & R_1^{\mathrm{T}}&
      \end{smallmatrix} \right| Q_1, Q_2, Q_3, Q_{B,1}, Q_{B,2},
    q, q \right)\times\\
  \times Z\left(\left.
      \begin{smallmatrix}
        &R_3^{\mathrm{T}} & R_2^{\mathrm{T}} & R_1^{\mathrm{T}}&\\
        \varnothing & & & &\varnothing\\
        &\varnothing &\varnothing &\varnothing &
      \end{smallmatrix} \right| \widetilde{Q}_1, \widetilde{Q}_2,
    \widetilde{Q}_3, Q_{B,1} Q_1 \widetilde{Q}_1^{-1}, Q_{B,2} Q_2
    \widetilde{Q}_2^{-1},
    q, q \right) = \\
  = \int_0^{\Lambda_1^{-1}} d^{N_1}_qx^{(1)} \int_0^{\Lambda_2^{-1}}
  d^{N_2}_qx^{(2)} \int_0^1 d^{N_3}_qx^{(3)}
  \mu_{\mathrm{DF}}(x^{(1)}, x^{(2)}, x^{(3)})
\end{multline}

\section{Refinement}
\label{sec:refin}
In this section we lift the previous results to the case of $t \neq
q$. In this setting topological string theory requires
\emph{refinement}. There exist two essentially equivalent forms of the
refined topological vertex, IKV~\cite{ref-tove} and
AK~\cite{ref-tove-ak}. The two versions are related by a simple change
of basis, so that the answer for any \emph{closed} string amplitude is
the same in both computations. However, open string amplitudes must be
transformed by matrices, attached to each external leg of the
diagram. We use IKV vertex throughout this paper since it seems to be
more convenient for comparison with the results in the unrefined case.

Refinement introduces a preferred direction and breaks the cyclic
(rotation) invariance of each individual topological vertex. However,
slicing invariance hypothesis states that the whole \emph{closed}
string amplitude remains invariant under rotations of the diagram, or
equivalently under the change of the preferred direction.

In this section we elucidate the mechanism of slicing invariance by
computing the partition function for vertical and horizontal
slicings. We set the preferred direction to be vertical. We observe
that while the horizontal slicing indeed reproduces the DF sum (and
the corresponding spectral dual Nekrasov
decomposition~\cite{spec-dual}), the vertical slicing \emph{does not}
give the factorized terms of the Nekrasov function. It requires a
further change of basis from Schur functions to \emph{generalized}
Macdonald polynomials. Moreover, unlike the change of basis, which
transforms the two types of refined topological vertices, this change
of basis is ``nonlocal'', i.e.\ it does not factorize into a product
of matrices each rotating its own external leg of the diagram. The
matrix depends on the whole array of states on the parallel external
legs of the diagram as well as on the distances (K\"ahler parameters)
between the legs.

The total matrix of the transformation is a certain triangular matrix
depending on the K\"ahler modulus for each external line. It is
natural to call this matrix generalized Kostka function by analogy
with the ordinary Kostka polynomials, which are the transition
coefficients between Schur and Macdonald polynomials. We therefore
explicitly identify the transformations corresponding to the change of
preferred direction. Algebraic meaning of these transformations will
be investigated elsewhere.

\subsection{$q$-Selberg measure}
\label{sec:df-integral}

Let us first write down the refinement of Eq.~\eqref{eq:3}, i.e.\
express the $q$-Sleberg measure for $t \neq q$ as a product of
Macdonald polynomials. We recall~\cite{genMD} that the Jackson
$q$-integral is in fact a sum with $x_i$ taking discrete values
$x_{R,i} = q^{R_i + 1} t^{N-i}$ where $R_i$ are the columns of a Young
diagram. The $q$-Selberg measure $\mu(x|u,v,N,q,t) = \Delta^{(q,t)}(x)
\prod_{i=1}^N \left( x_i^u \prod_{a=0}^{v-1} (1 - q^a x_i) \right)$
evaluated at $x_i(R)$ can be nicely expressed through Macdonald
polynomials:
\begin{multline}
  \label{eq:30}
  \frac{\mu(x_R|u,v,N,q,t)}{\mu(x_{\varnothing}|u,v,N,q,t)} =
  (-1)^{|R|} q^{(u+v+1)|R|} t^{(N-1)|R|} M_R^{(q,t)} \left( \frac{1 -
      t^{nN}}{1 - t^n} \right) M_{R^{\mathrm{T}}}^{(t,q)}
  \left( \frac{1 - t^{-n (N-1)} q^{- n (v+1)}}{1 - q^n} \right) =\\
  = (-1)^{|R|} q^{(u+v+1/2)|R|} t^{(N-3/2)|R|} M_R^{(q,t)} \left(p_n(
    t^{-\rho} ) - p_n (t^N t^{-\rho}) \right)
  M_{R^{\mathrm{T}}}^{(t,q)} \left( p_n (q^{-\rho}) - p_n (t^{1-N}
    q^{-v-1} q^{-\rho} ) \right) =\\
  = (-1)^{|R|} q^{u|R|} t^{N |R|} M_R^{(q,t)} \left( \frac{1 -
      q^{n(v+1)} t^{n(N-1)} }{1 - t^n} \right) M_{R^{\mathrm{T}}}^{(t,q)}
  \left( \frac{1 - t^{-n N}}{1 - q^n} \right)=\\
  = (-1)^{|R|} q^{(u-1/2)|R|} t^{(N-1/2)|R|} M_R^{(q,t)} \left(p_n(
    t^{-\rho} ) - p_n ( q^{v+1} t^{N-1} t^{-\rho}) \right)
  M_{R^{\mathrm{T}}}^{(t,q)} \left( p_n (q^{-\rho}) - p_n (t^{-N} q^{-\rho} ) \right).
\end{multline}
Notice the following useful symmetry
\begin{equation}
  \label{eq:50}
  \frac{\mu(x_R|u,v,N,q,t)}{\mu(x_{\varnothing}|u,v,N,q,t)} =
  \frac{\mu(x_R|u,-v-2+2\beta,N + (v + 1 - \beta)/
      \beta,q,t)}{\mu(x_{\varnothing}|u,-v-2+2\beta,N + (v + 1 - \beta)/
      \beta,q,t)}.
\end{equation}

We will use this form of the Selberg measure to identify the DF
integrals of the $q$-deformed CFT with the amplitudes of the
topological string on toric CY backgrounds.

\subsection{Generalized bifundamental kernel}
\label{sec:gener-bifund-kern}

Having understood the \emph{measure} of the $q$-Selberg integrals in
terms of Macdonald polynomials, we now proceed to describe \emph{what}
is being averaged in the Nekrasov decomposition of the conformal
block. Recall that to obtain the chain or quiver-like decomposition of
the block one needs to choose a special basis of intermediate states,
so that the matrix elements reproduce the individual terms of the
Nekrasov partition function. For the unrefined case this basis was
simply given by the product of Schur polynomials, and the matrix
elements were give by $q$-Selberg averages of two \emph{bifundamental
  kernels} $N_{AB}[x]$ (as in Eq.~\eqref{eq:15}). For the refined case
the basis is more elaborate: it is given by \emph{generalized}
Macdonald polynomials, which depend on two Young diagrams and do not
factorize into products of two polynomials. The relevant matrix
elements are given by the $q$-Selberg average of what we call
\emph{generalized} bifundamental kernel $\widetilde{N}_{AB,CD}$, a
convolution of two generalized Macdonald polynomials:
\begin{multline}
  \label{eq:66}
  \widetilde{N}^{(q,t)}_{AB,CD}(u,v,N|x) = \sum_{E,F} \left(
    \frac{t}{q} \right)^{|E|+|F|}
  \frac{1}{||M_E^{(q,t)}||^2 ||M_F^{(q,t)}||^2 } \times \\
  \times M^{*(q,t)}_{AB/EF} \left( q^{u+1} t^{-1} \left| - \left(
        \frac{t}{q} \right)^n p_{-n} - \frac{1 - (t/q)^n}{1 - t^n}
      q^{nv} , p_{-n} + \frac{(t/q)^n -
        q^{n v}}{1 - t^n} \right. \right)\times \\
  \times M^{(q,t)}_{CD/EF} \left( q^{u+v+1} t^{2N-1} \left| p_n , -p_n
      - \frac{1 - q^{-n v}}{1 - t^{-n}} \right.\right),
\end{multline}
where
\begin{equation}
  ||M_E^{(q,t)}||^2 = \frac{C_E(q,t)}{C'_E(q,t)} = \prod_{(i,j)
    \in E} \frac{1 - q^{E_i - j + 1} t^{E^{\mathrm{T}}_j - i}}{1 -
    q^{E_i - j} t^{E^{\mathrm{T}}_j - i + 1}}\notag
\end{equation}
is the norm of Macdonald polynomials and generalized skew Macdonald
polynomial is given by\footnote{One can ask why we ``subtract''
  \emph{ordinary} Macdonald polynomials from the generalized ones to
  obtain the skew polynomials. In fact Eq.~\eqref{eq:58} is
  independent of the concrete choice of ``subtracted'' polynomials as
  long as they constitute a complete system and the Cauchy identity
  holds. We will return to this issue in
  sec.~\ref{sec:gener-kostka-funct}.}
\begin{equation}
  M_{AB/EF}^{(q,t)}(Q|p_n
  ,\bar{p}_n) = M^{(q,t)}_E \left( n \frac{1 - q^n}{1 - t^n}
    \frac{\partial}{\partial p_n} \right) M^{(q,t)}_F \left( n \frac{1 -
      q^n}{1 - t^n} \frac{\partial}{\partial \bar{p}_n} \right)
  M_{AB}^{(q,t)}(Q|p_n ,\bar{p}).\notag
\end{equation}

One can immediately notice that for $t=q$
\begin{equation}
  \widetilde{N}^{(q,q)}_{AB,CD}(u,v,N|x) = N_{AC}[x]
N_{BD}\left[-p_n(x)-\frac{1-q^{nv}}{1-q^n}\right],\notag
\end{equation}
exactly reproducing the unrefined case~\eqref{eq:15}. Generalized
Macdonald polynomials are obtained from the kernel by forgetting about
one of the two pairs of Young diagrams:
\begin{gather}
  \label{eq:67}
  \widetilde{N}^{(q,t)}_{AB,\varnothing \varnothing}(u,v,N|x) =
  M^{*(q,t)}_{AB} \left( q^{u+1} t^{-1} \left| - \left( \frac{t}{q}
      \right)^n p_{-n} - \frac{1 - (t/q)^n}{1 - t^n} q^{nv} , p_{-n} +
      \frac{(t/q)^n -
        q^{n v}}{1 - t^n} \right. \right),\\
  \widetilde{N}^{(q,t)}_{\varnothing \varnothing,CD}(u,v,N|x) =
  M^{(q,t)}_{CD} \left( q^{u+v+1} t^{2N-1} \left| p_n , -p_n - \frac{1
        - q^{-n v}}{1 - t^{-n}} \right.\right)
\end{gather}
One can also get the product of two \emph{ordinary} Macdonald
polynomials by setting the two ``cross-wise'' diagrams to be empty:
\begin{equation}
  \label{eq:68}
  \widetilde{N}^{(q,t)}_{\varnothing B, C \varnothing }(u,v,N|x) =  M^{(q,t)}_{B} \left(  p_{-n} + \frac{(t/q)^n -
      q^{n v}}{1 - t^n} \right) M^{(q,t)}_{C} (p_n)
\end{equation}

The most remarkable property of the generalized bifundamental kernel
is that its $q$-Selberg average is factorized into a product of simple
monomials~\eqref{eq:58}. More concretely, it is given by the
bifundamental contribution to the Nekrasov function (hence the name of
the kernel). Schematically
\begin{equation}
  \label{eq:74}
  \left\langle \widetilde{N}^{(q,t)}_{AB, CD}(u,v,N|x) \right\rangle
  \sim  \frac{z_{\mathrm{bifund}}\left([A,B],[C,D]
    \right)}{z_{\mathrm{vect}}^{1/2}\left( A,B \right)
    z_{\mathrm{vect}}^{1/2} \left( D, C \right) }.
\end{equation}
Averages of this kind can be obtained by using the loop equations for
$q$-Selberg integral (or $(q,t)$-matrix model). The full form of the
average~\eqref{eq:74} and technical details are summarized in
Appendix~\ref{sec:loop-equat-matr}. Thus, we prove that generalized
bifundamental kernel is indeed the relevant object to be averaged to
get the chain-like decomposition of the DF integral. Let us now try to
find similar objects in refined topological strings.

\subsection{Vertical slicing}
\label{sec:refin-bifund-kern}

We start from the refinement of the basic building block, i.e. the
four-point amplitude~\eqref{eq:17} and set preferred direction to be
vertical. The general refined amplitude depending on four Young
diagrams is given by Eq.~\eqref{eq:47}. For our purposes we need the
following specializations:
\begin{multline}
  \label{eq:54}
    Z \left( \left.
    \begin{smallmatrix}
      & \varnothing& \\
      C^{\mathrm{T}} & & A^{\mathrm{T}}\\
      & R &
    \end{smallmatrix}
  \right|Q_2,q,t\right)
= \quad \parbox{2.8cm}{\includegraphics[width=2.8cm]{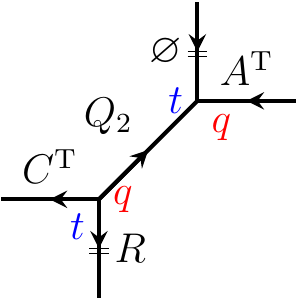}}\quad=\\
 = Z \left( \left.
    \begin{smallmatrix}
      & \varnothing & \\
      \varnothing & & \varnothing\\
      & \varnothing  &
    \end{smallmatrix}
  \right|Q_2,q,t\right) q^{\frac{||R||^2}{2}}
t^{-\frac{||R^{\mathrm{T}}||^2}{2}} \left( \frac{q}{t}
\right)^{\frac{|C|-|A|}{2}} (-Q_2)^{|C|} M_R^{(q,t)}
\left(p_n(t^{-\rho}) - p_n \left( Q_2 \sqrt{\frac{q}{t}}
    t^{-\rho} \right) \right) \times \\
\times \sum_E \chi_{C^{\mathrm{T}}/E} \left( - p_n(q^{-R} t^{-\rho}) +
  p_n \left( Q_2 \sqrt{\frac{q}{t}} t^{-\rho}\right) \right)
\chi_{A^{\mathrm{T}}/E} \left( \frac{1 - t^n}{1 - q^n} \left( p_n (q^R
    t^{\rho}) - \frac{q^n}{t^n} p_n \left( \sqrt{\frac{t}{q}} Q_2^{-1}
      t^{\rho} \right) \right) \right),
\end{multline}
and
\begin{multline}
  \label{eq:56}
    Z \left( \left.
    \begin{smallmatrix}
      & R & \\
      D^{\mathrm{T}} & & B^{\mathrm{T}}\\
      & \varnothing &
    \end{smallmatrix}
\right|Q_1,q,t\right) = \quad \parbox{2.8cm}{\includegraphics[width=2.8cm]{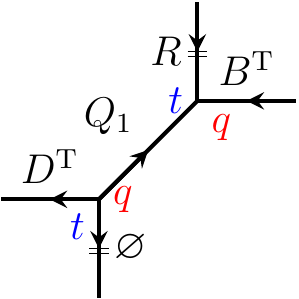}}\quad =\\
=Z \left( \left.
    \begin{smallmatrix}
      & \varnothing & \\
      \varnothing & & \varnothing\\
      & \varnothing  &
    \end{smallmatrix}
  \right|Q_1,q,t\right) q^{-\frac{||R||^2}{2}}
t^{\frac{||R^{\mathrm{T}}||^2}{2}} (-Q_1)^{|D|}
M_{R^{\mathrm{T}}}^{(t,q)} \left(p_n(q^{-\rho}) - p_n \left( Q_1
    \sqrt{\frac{t}{q}}
    q^{-\rho} \right) \right) \times \\
\times \sum_F \chi_{D^{\mathrm{T}}/F} \left( \frac{q^n}{t^n}
  p_n(q^{-R} t^{-\rho}) - p_n \left( Q_1^{-1} \sqrt{\frac{q}{t}}
    t^{-\rho}\right) \right) \chi_{B^{\mathrm{T}}/F} \left( -\frac{1 -
    t^n}{1 - q^n} \left( p_n (q^R t^{\rho}) - p_n \left( Q_1
      \sqrt{\frac{t}{q}} t^{\rho} \right) \right) \right),
\end{multline}

To make contact with DF integrals we make the identification
$\sqrt{\frac{t}{q}} Q_1 = t^{1-N} q^{-v-1}$, $\sqrt{\frac{q}{t}} Q_2 =
t^N$. We immediately notice that two Macdonald polynomials in
Eqs.~\eqref{eq:54} and~\eqref{eq:56} each give precisely one ``half''
of the $q$-Selberg measure~\eqref{eq:30}.

Skew Schur functions in Eqs.~\eqref{eq:54},~\eqref{eq:56} can be
rewritten through the discrete $q$-Selberg ``integration'' variables
$x_i(R) = q^{R_i + 1} t^{N-i}$ in the following way:
\begin{multline}
  \label{eq:26}
    Z \left( \left.
    \begin{smallmatrix}
      & \varnothing& \\
      C^{\mathrm{T}} & & A^{\mathrm{T}}\\
      & R &
    \end{smallmatrix}
\right|t^{N+\frac{1}{2}}q^{-\frac{1}{2}},q,t\right) = Z \left( \left.
    \begin{smallmatrix}
      & \varnothing & \\
      \varnothing & & \varnothing\\
      & \varnothing  &
    \end{smallmatrix}
  \right|t^{N+\frac{1}{2}}q^{-\frac{1}{2}},q,t\right)
(-1)^{|A|} \left( q^{-\frac{1}{2}} t^{-N} \right)^{|A|}  \left(
  q t^{2N-\frac{1}{2}}\right)^{|C|}\times\\
\times q^{\frac{||R||^2}{2}} t^{-\frac{||R^{\mathrm{T}}||^2}{2}} M_R^{(q,t)}
\left(p_n(t^{-\rho}) - p_n \left(t^N t^{-\rho} \right) \right)\times\\
\times \sum_E \left( \frac{t}{q} \right)^{|E|} \chi_{A/E} \left(
  \frac{1 - t^n}{1 - q^n} \left( - \left( \frac{t}{q} \right)^n p_n
    (x_i(R)) - \frac{1 - \left(\frac{q}{t}\right)^n}{1-t^{-n}} \right)
\right) \chi_{C/E} \left( p_{-n}(x_i(R)) \right),
\end{multline}
\begin{multline}
  \label{eq:27}
     Z \left( \left.
    \begin{smallmatrix}
      & R & \\
      D^{\mathrm{T}} & & B^{\mathrm{T}}\\
      & \varnothing &
    \end{smallmatrix}
\right|t^{\frac{1}{2}-N} q^{-v-\frac{1}{2}},q,t\right) = Z \left( \left.
    \begin{smallmatrix}
      & \varnothing & \\
      \varnothing & & \varnothing\\
      & \varnothing  &
    \end{smallmatrix}
  \right|t^{\frac{1}{2}-N} q^{-v-\frac{1}{2}},q,t\right)  (-1)^{|B|}
\left( q^{-1} t^{\frac{1}{2}-N} \right)^{|B|} \left(
  q^{-v+\frac{3}{2}}
  t^{-1} \right)^{|D|} \times\\
\times
q^{-\frac{||R||^2}{2}} t^{\frac{||R^{\mathrm{T}}||^2}{2}} M_{R^{\mathrm{T}}}^{(t,q)} \left(p_n(q^{-\rho}) - p_n
  \left(t^{1-N} q^{-1-v} q^{-\rho} \right) \right) \times\\
\times \sum_F \left(
  \frac{t}{q} \right)^{|F|} \chi_{B/F} \left( \frac{1 - t^n}{1 - q^n}
  \left( p_n (x_i(R)) + \frac{\left(\frac{q}{t}\right)^n - q^{-n v}}{1
      - t^{-n}} \right) \right) \chi_{D/E} \left( -p_{-n}(x_i(R)) -
  \left( \frac{t}{q} \right)^n \frac{1 - q^{n v}}{1 - t^n} \right).
\end{multline}
Let us glue two four-point functions~\eqref{eq:54} and~\eqref{eq:56}
to obtain the $q$-Selberg average:
\begin{multline}
  \label{eq:62}
Z \left( \left.
    \begin{smallmatrix}
      & \varnothing& \\
      C^{\mathrm{T}} & & A^{\mathrm{T}}\\
      D^{\mathrm{T}} & & B^{\mathrm{T}}\\
      & \varnothing &
    \end{smallmatrix}
  \right|Q_1,Q_2,Q_F,q,t\right)
=\quad \parbox{3.5cm}{\includegraphics[width=3.5cm]{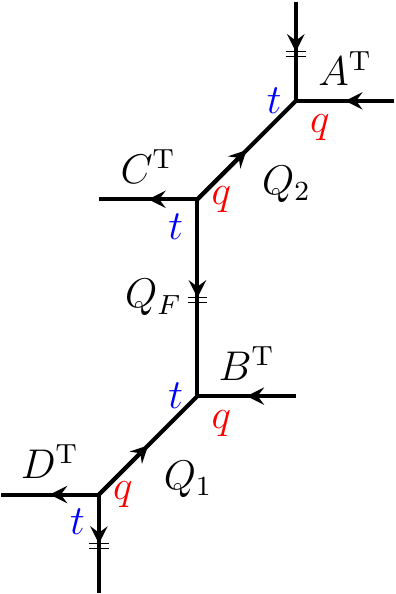}}
=\\
= \sum_R (-Q_F)^{|R|} Z \left( \left.
    \begin{smallmatrix}
      & \varnothing& \\
      C^{\mathrm{T}} & & A^{\mathrm{T}}\\
      & R &
    \end{smallmatrix}
\right|t^{N+\frac{1}{2}}q^{-\frac{1}{2}},q,t\right) Z \left( \left.
    \begin{smallmatrix}
      & R & \\
      D^{\mathrm{T}} & & B^{\mathrm{T}}\\
      & \varnothing &
    \end{smallmatrix}
  \right|t^{\frac{1}{2}-N} q^{-v-\frac{1}{2}},q,t\right) = \\
= \frac{S_{u,v,N,q,t}\, Z \left( \left.
    \begin{smallmatrix}
      & \varnothing & \\
      \varnothing & & \varnothing\\
      & \varnothing  &
    \end{smallmatrix}
  \right|t^{N+\frac{1}{2}}q^{-\frac{1}{2}},q,t\right)
Z \left( \left.
    \begin{smallmatrix}
      & \varnothing & \\
      \varnothing & & \varnothing\\
      & \varnothing  &
    \end{smallmatrix}
  \right|t^{\frac{1}{2}-N}
  q^{-v-\frac{1}{2}},q,t\right)}{\mu(x_{\varnothing}|u,v,N,q,t)} \times\\
\times (-1)^{|A|+|B|} \left( q^{-\frac{1}{2}} t^{-N} \right)^{|A|}
\left( q t^{2N-\frac{1}{2}}\right)^{|C|} \left( q^{-1}
  t^{\frac{1}{2}-N}
\right)^{|B|} \left( q^{-v+\frac{3}{2}} t^{-1} \right)^{|D|}\times\\
\times \Biggl\langle\sum_{E,F} \left( \frac{t}{q}
\right)^{|E|+|F|} \chi_{A/E} \left( \frac{1 - t^n}{1 - q^n} \left( -
    \left( \frac{t}{q} \right)^n p_n (x_{R,i}) - \frac{1 -
      \left(\frac{q}{t}\right)^n}{1-t^{-n}} \right)
\right) \chi_{C/E} \left( p_{-n}(x_{R,i}) \right) \times\\
\times \chi_{B/F} \left( \frac{1 - t^n}{1 - q^n} \left( p_n (x_{R,i}) +
    \frac{\left(\frac{q}{t}\right)^n - q^{-n v}}{1 - t^{-n}} \right)
\right) \chi_{D/E} \left( -p_{-n}(x_{R,i}) - \left( \frac{t}{q}
  \right)^n \frac{1 - q^{n v}}{1 - t^n} \right) \Biggr\rangle ,
\end{multline}
where $S_{u,v,N,q,t}$ is the $q$-Selberg integral without insertions.
We have used the identity~\eqref{eq:30} and made the identification
$Q_F = q^{u+v+\frac12} t^{N-\frac32}$. At this point one observes that
the expression under the average \emph{is not} the generalized
bifundamental kernel~\eqref{eq:66}, which would give the bifundamental
contribution as an average. Instead it is simply a product of two
Schur polynomials. However, closer look reveals that the
\emph{arguments} of the Schur polynomials exactly match (up to the
factors $\frac{1-t^n}{1-q^n}$) the arguments of the generalized
bifundamental kernel after the elementary transformation $x_i \to
q^{1-v} x_i^{-1}$ as can be seen e.g.\ from Eq.~\eqref{eq:60}. This
leads us to the relation between Nekrasov functions and the vertical
slicing of the refined amplitude. To get it we will need to introduce
generalized Kostka functions.

\subsection{Generalized Kostka functions}
\label{sec:gener-kostka-funct}

The basis of generalized Macdonald polynomials can be reexpanded in
terms of Schur polynomials:
\begin{gather}
  \label{eq:69}
  M^{(q,t)}_{AB}(Q|p_n, s_n) = \sum_{C,D} K_{AB}^{CD} (Q,q,t)
  \chi_C(p_n) \chi_D(s_n),\\
  M^{*(q,t)}_{AB}(Q|p_n, s_n) =  ||M_A^{(q,t)}||^2 ||M_B^{(q,t)}||^2 \sum_{C,D} K_{AB}^{*CD} (Q,q,t)
  \chi_C\left(\frac{1-q^n}{1-t^n} p_n \right) \chi_D\left(\frac{1-q^n}{1-t^n} s_n \right), \label{eq:73}
\end{gather}
where the coefficients $K_{CD}^{AB}(Q,q,t)$ can naturally be called
\emph{generalized Kostka functions} by analogy with the ordinary
Kostka polynomials\footnote{Our definition of generalized Kostka
  functions can be modified slightly to turn them into
  polynomials. This is achieved by using a different normalization of
  generalized Macdonald polynomials, called
  $\widetilde{M}^{(q,t)}_{AB}$ in Eqs.~(19), (20) in~\cite{genMD}.},
defined as
\begin{equation}
  \label{eq:70}
  M_A^{(q,t)}(p_n) = \sum_B K_A^B(q,t) \chi_B(p_n).
\end{equation}
As an explicit example we give here generalized Kostka functions for
the first level:
\begin{gather}
  \label{eq:71}
 \left. K_{AB}^{CD}(Q,q,t)\right|_{|A|+|B|=|C|+|D|=1} = \left(
\begin{array}{cc}
 1 & 0 \\
 \frac{1- \frac{t}{q}}{1 - Q} & 1 \\
\end{array}
\right),\\
\left. K_{AB}^{*CD}(Q,q,t)\right|_{|A|+|B|=|C|+|D|=1} = \left(
\begin{array}{cc}
  1 & -\frac{1-\frac{t}{q}}{(1-Q)} \\
  0 & 1 \\
\end{array}
\right).\label{eq:79}
\end{gather}

To transform skew Schur into skew generalized Macdonald polynomials in
the generalized bifundamental kernel we use the following identity,
which is the consequence of the Cauchy completeness theorem:
\begin{multline}
  \label{eq:55}
  \sum_E \chi_{A/E}(p_n) \chi_{B/E}\left( \frac{1-t^n}{1-q^n} r_n
  \right) = \exp \left( \sum_{n \geq 1} \frac{1}{n} \frac{1-q^n}{1 -
      t^n} \frac{\partial}{\partial p_n} \frac{\partial}{\partial r_n}
  \right) \chi_A(p_n) \chi_B\left( \frac{1-t^n}{1-q^n} r_n
  \right) =\\
  = \sum_E \left\{ M^{(q,t)}_E \left( n \frac{1-q^n}{1-t^n}
      \frac{\partial}{\partial p_n} \right) \chi_A(p_n) \right\}
  \left\{ M^{(q,t)}_E \left( n \frac{1-q^n}{1-t^n}
      \frac{\partial}{\partial r_n} \right) \chi_B\left(
      \frac{1-t^n}{1-q^n} r_n \right)\right\}.
\end{multline}
The combination of Macdonald functions of the conjugated power sums
$p_n$ in the last line exactly reproduces that in the generalized
bifundamental kernel~\eqref{eq:66}. What is left is to transform Schur
polynomials into generalized Macdonald ones with the help of
generalized Kostka functions~\eqref{eq:69},~\eqref{eq:73}.

Eventually, we obtain the connection between refined topological
string amplitude with vertical slicing and bifundamental Nekrasov
function:
\begin{multline}
  \label{eq:72}
  \sum_{A,B,C,D} K^{CD}_{W_1 W_2}\left( (q Q_F Q_1)^{-1} , q,t \right)  Z \left( \left.
    \begin{smallmatrix}
      & \varnothing& \\
      C^{\mathrm{T}} & & A^{\mathrm{T}}\\
      D^{\mathrm{T}} & & B^{\mathrm{T}}\\
      & \varnothing &
    \end{smallmatrix}
  \right|Q_1,Q_2,Q_F,q,t\right) K^{*AB}_{Y_1 Y_2} \left( (q
  Q_F Q_2)^{-1}, q, t \right) =\\
=\left[
  \begin{smallmatrix}
    Q_1 = t^{\frac{1}{2}-N}q^{-v-\frac{1}{2}}\\
    Q_2 = t^{N+\frac{1}{2}} q^{-\frac{1}{2}}\\
    Q_F = q^{u+v+\frac12}t^{N-\frac32}
  \end{smallmatrix}
\right]=\\
=\frac{S_{u,v,N,q,t}\, Z \left( \left.
    \begin{smallmatrix}
      & \varnothing & \\
      \varnothing & & \varnothing\\
      & \varnothing  &
    \end{smallmatrix}
  \right|t^{N+\frac{1}{2}}q^{-\frac{1}{2}},q,t\right)
Z \left( \left.
    \begin{smallmatrix}
      & \varnothing & \\
      \varnothing & & \varnothing\\
      & \varnothing  &
    \end{smallmatrix}
  \right|t^{\frac{1}{2}-N}
  q^{-v-\frac{1}{2}},q,t\right)}{\mu(x(\varnothing)|u,v,N,q,t)} \times\\
\times (-1)^{|A|+|B|} \left( q^{v-\frac{3}{2}} t^{-N} \right)^{|A|}
\left( q^v t^{2N-\frac{1}{2}}\right)^{|C|} \left( q^{v-2}
  t^{\frac{1}{2}-N}
\right)^{|B|} \left( q^{\frac{1}{2}} t^{-1} \right)^{|D|}\times\\
\times \left \langle \widetilde{N}_{AB,CD}(u,v,N|q^{1-v} x_R^{-1}) \right\rangle_{u,v,N,q,t} =\\
=(-1)^{|B|+|C|} q^{-(1+v) |A| - 2 |C| - (u+2v+1) |B| +(u-2)|D|}
t^{|A|+(1-2N)|B| + |C| -
  (4N + 1)|D|}\times\\
\times t^{\sum_{(i,j)\in A} i + 2\sum_{(i,j)\in B} i - \sum_{(i,j)\in
    D} i} q^{ - \sum_{(i,j)\in B} j +
  \sum_{(i,j)\in C} j + 2\sum_{(i,j)\in D} j} \times \\
\times
\frac{z_{\mathrm{bifund}}^{(q,t)}\left([A,B],[C,D],\frac{-u-v-1-2\beta
      N + \beta}{2} , \frac{-u-1+\beta}{2}, -\frac{v}{2}-1+ \beta
  \right)}{C'_A(q,t) C'_B(q,t) C'_C(q,t) C'_D(q,t) G_{AB}^{(q,t)}
  (q^{-u-v-1}t^{1-2N}) G_{DC}^{(q,t)} (q^{u+1}t^{-1}) }.
\end{multline}
Note that generalized Kostka functions in this formula depend on the
``distance'' (in the sense of K\"ahler parameters) between the pairs
of horizontal external legs of the toric diagram. Let us also point
out that our Kostka functions are $q$-deformation of the coefficients
of the abelianization map acting on the instanton moduli space.

\subsection{Horizontal slicing. DF representation and spectral dual
  Nekrasov function.}
\label{sec:horiz-slic-df}

Let us glue three pieces \eqref{eq:54} together horizontally to obtain
the DF integrand for five-point conformal block and its AGT dual ---
$U(2)^2$ quiver gauge theory. The resulting amplitude is equal to
``half'' of the total DF measure
$\mu_{\mathrm{DF}}(x_{(1)},x_{(2)},x_{(3)})$ evaluated at discrete
points $x_{a,i} = q^{R_{a,i} + 1}t^{N_a - i}$. We get:
\begin{multline}
  \label{eq:28}
  Z\left(\left.
      \begin{smallmatrix}
        &\varnothing & \varnothing & \varnothing&\\
        \varnothing & & & &\varnothing\\
        &R_3 & R_2 & R_1&
      \end{smallmatrix} \right| Q_1, Q_2, Q_3, Q_{B,1}, Q_{B,2},
    q, t \right) = \quad \parbox{7.5cm}{\includegraphics[width=7.5cm]{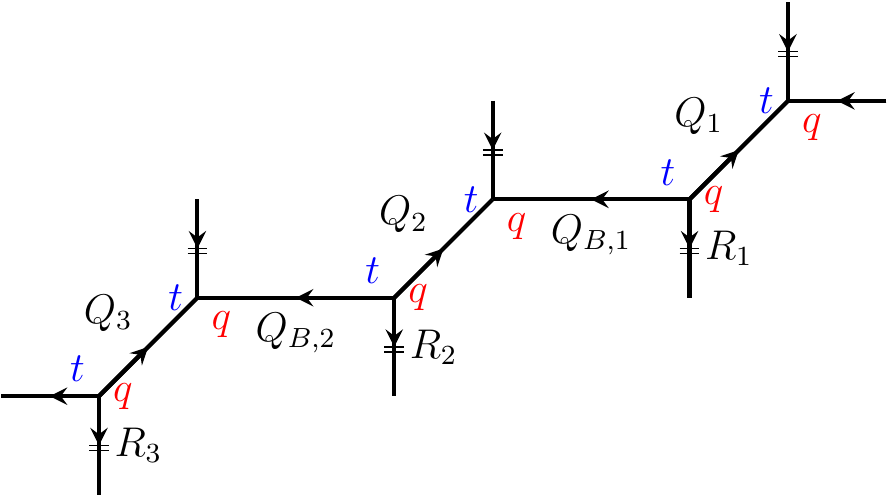}} \quad =\\
  = \prod_{a=1}^3 \left[Z \left( \left.
    \begin{smallmatrix}
      & \varnothing & \\
      \varnothing & & \varnothing\\
      & \varnothing  &
    \end{smallmatrix}
  \right|t^{N_a +\frac{1}{2}}q^{-\frac{1}{2}},q,t\right)
q^{\frac{||R_a||^2}{2}} t^{-\frac{||R_a^{\mathrm{T}}||^2}{2}}
M_{R_a}^{(q,t)} \left(p_n(t^{-\rho}) - p_n \left(t^{N_{a}} t^{-\rho}
  \right) \right)\right] \times\\
\times \prod_{m=0}^{\beta-1} \left[ \prod_{i=1}^{N_1}
  \prod_{j=1}^{N_2} \left( 1 - \Lambda_1 q^m \frac{x_{R_2, j}}{x_{R_1,
        i}} \right) \prod_{j=1}^{N_2} \prod_{k=1}^{N_3} \left( 1 -
    \Lambda_2 q^m \frac{x_{R_3, k}}{x_{R_2, j}} \right)
  \prod_{i=1}^{N_1} \prod_{k=1}^{N_3} \left( 1 - \Lambda_1 \Lambda_2
    q^m
    \frac{x_{R_3, k}}{x_{R_1, i}} \right)  \right]\times\\
\times \prod_{m=0}^{\beta-2} \left[ \prod_{i=1}^{N_1} \left( 1 -
    \frac{\Lambda_1 q^{-m}}{x_{R_1, i}} \right)^{-1} \left( 1 -
    \frac{\Lambda_1 \Lambda_2 q^{-m}}{x_{R_1, i}} \right)^{-1}
  \prod_{j=1}^{N_2} \left( 1 - \frac{\Lambda_2 q^{-m}}{x_{R_2, j}}
  \right)^{-1} \right]
\end{multline}
Three Macdonald polynomials in the second line can be thought of as a
``half'' of the three $q$-Selberg measures, corresponding to three
integration contours in the DF representation.

Moreover, the measure~\eqref{eq:28} can be evaluated explicitly and
also gives the ``half'' of the spectral dual Nekrasov function with
gauge group $U(3)$, cf.~\eqref{eq:43} (the other half of the factors
comes from the lower half of the diagram):
\begin{multline}
  \label{eq:78}
  Z\left(\left.
      \begin{smallmatrix}
        &\varnothing & \varnothing & \varnothing&\\
        \varnothing & & & &\varnothing\\
        &R_3 & R_2 & R_1&
      \end{smallmatrix} \right| Q_1, Q_2, Q_3, Q_{B,1}, Q_{B,2}, q, t
  \right) =\\
  = \frac{q^{\frac{-||R_1||^2 + ||R_3||^2}{2}}
    t^{||R_1^{\mathrm{T}}||^2 + \frac{||R_2^{\mathrm{T}}||^2}{2}}
    Z_{\varnothing}(Q_1) Z_{\varnothing}(Q_2) Z_{\varnothing}(Q_3)
    Z_{\varnothing}(Q_{B,2}) Z_{\varnothing}(Q_{B,2} Q_2 Q_3)
    Z_{\varnothing}(Q_{B,1}) }{C_{R_1}'(q,t) C_{R_2}'(q,t)
    C_{R_3}'(q,t) Z_{\varnothing}\left(\sqrt{\frac{q}{t}} Q_{B,2} Q_2
    \right) Z_{\varnothing}\left(\sqrt{\frac{t}{q}} Q_{B,2} Q_3
    \right) Z_{\varnothing}\left(\sqrt{\frac{q}{t}} Q_1 Q_{B,2}\right)} \times\\
  \times \frac{(Q_{B,1}^2 Q_{B,2} Q_2)^{|R_1|} (-Q_{B,2})^{|R_2|}
    Z_{\varnothing}(Q_1 Q_2 Q_{B,1}) Z_{\varnothing}(Q_{B,1} Q_{B,2}
    Q_2) Z_{\varnothing}(Q_1 Q_2 Q_3 Q_{B,1}
    Q_{B,2})}{Z_{\varnothing}\left(\sqrt{\frac{t}{q}} Q_2
      Q_{B,1}\right) Z_{\varnothing}\left(\sqrt{\frac{q}{t}} Q_1 Q_2
      Q_{B,1} Q_{B,2}\right) Z_{\varnothing}\left(\sqrt{\frac{t}{q}}
      Q_{B,1} Q_{B,2} Q_2 Q_3 \right)}\times\\
  \times G^{(q,t)}_{R_1 \varnothing} \left( \sqrt{\frac{q}{t}} Q_1
  \right) G^{(q,t)}_{R_1 \varnothing} \left( \sqrt{\frac{q}{t}}
    Q_{B,1}^{-1} \right) G^{(q,t)}_{R_1 \varnothing} \left(
    \sqrt{\frac{q}{t}}
    (Q_{B,1} Q_{B,2} Q_2)^{-1} \right) \times\\
  \times G^{(q,t)}_{R_2 \varnothing} \left( \sqrt{\frac{q}{t}} Q_2
  \right) G^{(q,t)}_{R_2 \varnothing} \left( \sqrt{\frac{q}{t}}
    Q_{B,2}^{-1} \right) G^{(q,t)}_{R_2 \varnothing} \left(
    \sqrt{\frac{q}{t}} Q_1 Q_2 Q_{B,1} \right)\times\\
  \times \frac{G^{(q,t)}_{R_3 \varnothing} \left( \sqrt{\frac{q}{t}} Q_3
  \right) G^{(q,t)}_{R_3 \varnothing} \left( \sqrt{\frac{q}{t}}
    Q_{B,2} Q_2 Q_3 \right) G^{(q,t)}_{R_3 \varnothing} \left(
    \sqrt{\frac{q}{t}} Q_1 Q_2 Q_3 Q_{B,1} Q_{B,2}
  \right)}{G^{(q,t)}_{R_3 R_2} (Q_{B,1} Q_3) G^{(q,t)}_{R_2 R_1}
  (Q_2 Q_{B,1}) G^{(q,t)}_{R_3 R_1} (Q_{B,1} Q_{B,2} Q_2 Q_3)},
\end{multline}
where $Z_{\varnothing}(Q) = Z \left( \left.
    \begin{smallmatrix}
      & \varnothing & \\
      \varnothing & & \varnothing\\
      & \varnothing &
    \end{smallmatrix}
  \right|Q,q,t\right)$. This is a manifestation of the spectral
duality for Nekrasov functions~\cite{spec-dual}: while vertical
slicing of the toric diagram gives Nekrasov function for $U(2)^3$
quiver gauge theory, the horizontal slicing yields its spectral dual
--- gauge theory with a single $U(3)$ gauge group. In the language of
conformal blocks~\cite{genMD} this means that \emph{both} the Jackson
integral and the sum over complete basis of generalized Macdonal
polynomials have the form of Nekrasov decompositions, which are
spectral dual to each other. For refined topological strings only one
Nekrasov decomposition can be obtained for a given choice of preferred
direction --- the cut should dissect the preferred edges. If one cuts
along a different direction the amplitudes \emph{do not} reproduce the
Nekrasov functions, as can be seen in Eq.~\eqref{eq:62}. However,
there is still a way to see the dual decomposition: preferred
direction can be changed with the help of generalized Kostka
functions~\eqref{eq:71},~\eqref{eq:79}.

\section{Conclusions and discussion}
\label{sec:discussion}
We have investigated the connection between $q$-deformed conformal
blocks and topological strings. This connection arises in the
following way. Due to the AGT relation conformal blocks are equal to
Nekrasov partition functions, which can be obtained by the geometric
engineering technique, as compactifications of type IIA strings (or,
more generally, M-theory) on toric CY threefold. String partition
function on the threefold is equal to partition function of
topological strings.

We obtain an explicit dictionary between the objects in CFT and
elements of the corresponding toric diagram, summarized in
Table~\ref{tab:1}. For the case of $t=q$ we introduce the
bifundamental kernel~\eqref{eq:35}, compute its $q$-Selberg
averages~\eqref{eq:40} and show that they reproduce Nekrasov partition
function. We also study spectral duality of conformal blocks and
generalize the statements of~\cite{genMD} to multipoint blocks. Most
importantly, we study the ever-troublesome case of $t \neq q$, where
we introduce \emph{generalized} bifundamental kernel~\eqref{eq:66}. We
compute the average of the generalized kernel --- it satisfies the
most general of all the so far encountered ``factorization of
averages'' type identities~\eqref{eq:58} --- and is again given by
Nekrasov function. We interpret the change of preferred direction of
refined topological strings as a change of basis between generalized
Macdonald and Schur polynomials, which is performed by generalized
Kostka functions~\eqref{eq:69},~\eqref{eq:73}.

\begin{table}[h]
  \centering
  \setlength{\tabcolsep}{8pt}
  \begin{tabular}{|c|c|}
    \hline
    \textbf{CFT} & \textbf{Topological string}\\
    \hline
    Conformal block $\mathcal{B}_k$, Fig.~\ref{fig:8} & Closed string amplitude on toric
    strip CY $Z_{\mathrm{top}}$, Fig.~\ref{fig:1}\\
    \hline
    $q$-Selberg measure~\eqref{eq:30} & Two four-point conifold amplitudes~\eqref{eq:17},~\eqref{eq:18}\\
    \hline
    DF integral~\eqref{eq:33} & Horizontal slicing, vertical preferred
    direction~\eqref{eq:53},~\eqref{eq:28}\\
    \hline
    Nekrasov/generalized Macdonald decomposition~\eqref{eq:29} & Vertical slicing, horizontal preferred
    direction~\eqref{eq:78}\\
    \hline
    Decomposition in Schur polynomials & Vertical slicing, vertical
    preferred direction~\eqref{eq:62}\\
    \hline
    Rotation of preferred direction by $\frac{\pi}{2}$ & Generalized Kostka function~\eqref{eq:71},~\eqref{eq:79}\\
    \hline
  \end{tabular}
  \caption{Summary of the CFT/topological string dictionary.}
  \label{tab:1}
\end{table}

Of course the expansion we have considered is not limited to the case
of $U(2)$ gauge theories and Virasoro conformal blocks. $U(N)$/$W_N$
story goes along the same lines. In this setting $q$-Selberg integrals
are replaced by the $A_N$ $q$-Selberg integrals, their measure being
given by the product of several basic building
blocks~\eqref{eq:47}. Generalized Macdonald polynomials, bifundamental
kernels and Kostka functions can also be found for the $U(N)$ case.

It would be extremely interesting to understand the relation of the
character/topological string decomposition of conformal blocks from
the point of view of Seiberg-Witten integrable systems. One relation
is provided by the quantum spectral curve for DF integrals
\cite{qt-spec-curve}, which in Nekrasov-Shatashvili limit reproduces
the quantum spectral curve (Baxter TQ equation) of the relevant
Seiberg-Witten system, the XXZ spin chain. Of course, a general method
to obtain quantum spectral curves from the toric data is
desirable. Let us also mention that in this way one can study the
mirror symmetry between the B-model CY, encoded in the spectral curve
and the A-side toric CY described by the topological vertex formalism.

In the four dimensional limit generalized Kostka polynomials coincide
with the coefficients of the abelianization map acting on the fixed
points in the cohomology of the instanton moduli space. Explicit
combinatorial expressions for these coefficients were obtained
in~\cite{Smirnov-abel}. It would be interesting to understand these
formulas from the point of view of refined topological strings.

The product of generalized Kostka matrices turns out to be an
interesting algebraic object. We can reason in the following way. Let
us first expand generalized Macdonald polynomials in terms of products
of Schur polynomials using the Kostka matrix. Then we exchange the two
Schur polynomials and apply the reverse Kostka transformation. Thus we
obtain another set of generalized Macdonald polynomials. However the
two sets are clearly related. Recall that generalized Macdonald
polynomials are eigenfunctions of the operator $H_1^{\mathrm{gen}} =
\Delta(H_1)$, which is given by the Ding-Iohara coproduct, acting on
trigonometric Ruijsenaars Hamiltonian $H_1$. The second set of
generalized Macdonald polynomials is obtained by acting on the same
Ruijsenaars Hamiltonian with the opposite coproduct
$\Delta^{\mathrm{op}}$. As in any quasitriangular Hopf algebra, there
is an $R$-matrix performing the transformation from one coproduct to
the other. The two sets of generalized Macdonald polynomials are
therefore also related by the same $R$-matrix. This is the
$K$-theoretic version of the instanton $R$-matrix~\cite{Smirnov-R}
with spectral parameter being the parameter of generalized Macdonald
polynomials. The implications of this observation and the relation
between toric CY and integrable systems will be studied elsewhere.

\section*{Acknowledgements}

We thank Andrey Smirnov for clarifying the concept of instanton
$R$-matrix to us. Our work is partly supported by grants
NSh-1500.2014.2 and 15-31-20832-Mol-a-ved, by RFBR grants 13-02-00478,
by joint grants 15-52-50034-YaF, 15-51-52031-NSC-a, by
14-01-92691-Ind-a and by the Brazilian National Counsel of Scientific
and Technological Development. Y.Z.\ is supported by the ``Dynasty''
foundation stipend.

\appendix

\section{Five-dimensional Nekrasov functions and AGT relations}
\label{sec:five-dimens-nekr}
The Nekrasov partition function for the $U(N)$ theory with $N_f = 2N$
fundamental hypermultiplets is given by
\begin{equation}
\label{eq:43}
Z_{\mathrm{Nek}}^{5d,\, U(N)} = \sum_{\vec{A}} \Lambda^{|\vec{A}|}
\frac{z_{\mathrm{fund}} (\vec{A}, \vec{m}^{+}, \vec{a})
  z_{\overline{\mathrm{fund}}} (\vec{A}, \vec{m}^{-}, \vec{a})
}{z_{\mathrm{vect}}(\vec{A},\vec{a})} = \sum_{\vec{A}} \Lambda^{|\vec{A}|}
\frac{\prod_{i=1}^{N}\prod_{f=1}^{N} f_{A_i}^{+} (q^{m_f^{+} + a_i}) f_{A_i}^{-} (q^{m_f^{-} + a_i})
}{z_{\mathrm{vect}}(\vec{A},\vec{a})}\,,
\end{equation}
where $f_A^{\pm} (q^x) = \prod_{(i,j) \in A} \left(1 - q^{\pm x}
  t^{\pm ( i - 1)} q^{\mp (j - 1 )} \right)$,
$z_{\mathrm{vect}}(\vec{A},\vec{a}) = \prod_{i,j=1}^{N} G^{(q,t)}_{A_i
  A_j}(q^{a_i - a_j})$ and
\begin{multline}
\label{eq:42}
  G^{(q,t)}_{AB} (q^x)= \prod_{(i, j) \in A} \left( 1 - q^x q^{A_i - j}
    t^{B^{\mathrm{T}}_j - i + 1} \right) \prod_{(i,j) \in B}\left(1 -
    q^x q^{-B_i + j - 1} t^{-A^{\mathrm{T}}_j + i} \right) =\\
 = \prod_{(i, j) \in B} \left( 1 - q^x q^{A_i - j}
    t^{B^{\mathrm{T}}_j - i + 1} \right) \prod_{(i,j) \in A}\left(1 -
    q^x q^{-B_i + j - 1} t^{-A^{\mathrm{T}}_j + i} \right),
\end{multline}
in particular
\begin{gather}
  \label{eq:76}
  G^{(q,t)}_{A \varnothing} (q^x) = \prod_{(i,j)\in A} \left( 1 - q^x
    q^{j-1} t^{1-i} \right) = f^{-}_A(q^{-x}),\\
  G^{(q,t)}_{\varnothing A} (q^x) = \prod_{(i,j)\in A} \left( 1 - q^x
    q^{1-j} t^{i-1} \right) = f^{+}_A(q^x),
\end{gather}

We will write $a$ instead of $\vec{a}=(a,-a)$ for $N=2$. The AGT
relations for $N=2$ are:
\begin{gather}
  u_{+} = m_1^{+} - m_2^{+} - 1 + \beta\,, \qquad \qquad u_{-} = -1 +
  \beta -2a \,, \notag\\
  v_{+} = - m^{+}_1 - m^{+}_2 \,, \qquad \qquad v_{-} =
  - m_1^{-} - m_2^{-}\,, \label{eq:41}\\
  \beta n_{+} = -a + m^{+}_2\,, \qquad \qquad \beta n_{-} = a +
  m^{-}_2\,,\notag
\end{gather}
where $a_1 = - a_2 = a$. Masses $m_a$, vevs $a_i$, radius $R_5$ of the
fifth dimension and $\epsilon_{1,2}$ all have dimensions of mass. In
this paper we set the overall mass scale so that $\epsilon_1 = -b^2$,
$\epsilon_2 = 1$ and $q = e^{-R_5}$. The $t$ parameter in Macdonald
polynomials is related to $q$ by $t = q^{\beta}$ with $\beta = b^2$.

More generally, one can consider quiver gauge theories with gauge
groups $U(N)^k$ and bifundamental matter hypermultiplets as shown in
Fig.~\ref{fig:2}. The corresponding Nekrasov function is
\begin{multline}
  \label{eq:61}
  Z_{\mathrm{Nek}}^{5d,\, U(N)^k} = \sum_{\vec{Y}_a} \Lambda_1^{|\vec{Y}_1|} \cdots
  \Lambda_k^{|\vec{Y}_k|} \prod_{f=1}^N \prod_{i=1}^N
  f^{+}_{Y_{1,i}}\left(q^{m_f^{+} + a_{1,i}} \right)
  \frac{1}{z_{\mathrm{vec}}(\vec{Y}_1, \vec{a}_1)}
  z_{\mathrm{bifund}}\left(\vec{Y}_1, \vec{Y}_2, \vec{a}_1, \vec{a}_2,
  m_{\mathrm{bifund},1}\right) \cdots\\
  \cdots z_{\mathrm{bifund}}\left(\vec{Y}_{k-1}, \vec{Y}_k, \vec{a}_{k-1},
  \vec{a}_k, m_{\mathrm{bifund},k-1}\right)
  \frac{1}{z_{\mathrm{vec}}(\vec{Y}_k , \vec{a}_k)} \prod_{f=1}^N
  \prod_{i=1}^N f^{-}_{Y_{k,i}}\left(q^{m_f^{-} + a_{k,i}} \right)
\end{multline}
where the bifundamental contribution is given by
$z_{\mathrm{bifund}}(\vec{Y}, \vec{W}, \vec{a}, \vec{b}, m) =
\prod_{i=1}^N \prod_{j=1}^N G_{Y_i W_j}^{(q,t)} \left( q^{a_i - b_j -
    m} \right)$.

\section{Loop equations for matrix elements}
\label{sec:loop-equat-matr}

We would like to compute the $q$-Selberg average of a function
$f(p_n)$ which is polynomial both in $p_n$ and $p_{-n}$. To do this
we use an improved version of the loop equations~\cite{betadefo}:
\begin{multline}
  \label{eq:19}
  \Biggl\langle \left( \frac{1}{q} - \frac{1}{z} \right) f(p_n) \exp
  \left( \sum_{n
      \geq 1} \frac{1 - t^n}{n} z^{-n} p_n \right) + \\
  +t^{2N-1} q^{u+1} \left( q^{v-1} - \frac{1}{z} \right) \Biggl\langle
  f(p_n + (1 - q^{-n}) z^n ) \exp \left( \sum_{n \geq 1} \frac{1 -
      t^{-n}}{n} q^n z^{-n} p_n \right) -\\
  - \mathrm{Res}_{\xi = 0} \left[\frac{t^{N-1} q^{u+1} (q^v \xi - 1)}{\xi (z
    - q \xi)} f(p_n + (q^n - 1) \xi^n) \exp \left( \sum_{n \geq 1} \frac{1 -
      t^n}{n} \xi^n p_{-n} \right)\right] - \frac{t^N f(p_n)}{z} +\\
  + \mathrm{Res}_{\xi = 0} \left[\frac{t^{2N-1} q^{u+1} (q^v - \xi)}{\xi (z \xi
    - q)} f(p_n + (q^n - 1) \xi^{-n}) \exp \left( \sum_{n \geq 1} \frac{1 -
      t^{-n}}{n} \xi^n p_n \right)\right] - \frac{f(p_n)}{q}\Biggr\rangle = 0
\end{multline}
Let us note that the expansion in positive and negative powers of $z$
lead to the same recurrence relations as it should. In addition to the
usual factorized formulas for the generalized Macdonald polynomials
these equations give the averages of the \emph{products} of two
Macdonald polynomials, one in $p_n$ the other in $p_{-n}$, e.g.:
\begin{equation}
  \label{eq:20}
  \left\langle \left( p_{-1} + \frac{t/q - q^v}{1 - t} \right)
    p_1\right\rangle = -\frac{t^{-N+1} (1 - t^N) (1 - q^u t^{N-1}) (1 -
    q^{1+u} t^N) (1 - q^{1+v} t^{N-1})}{(1 - q^u) (1-t)^2(1 - q^{2 + u + v} t^{2N-2})}.
\end{equation}

More generally
\begin{multline}
  \label{eq:22}
  \left\langle M_A \left( p_{-n} + \frac{t^n/q^n - q^{n v}}{1 - t^n}
    \right) M_B (p_n) \right\rangle =\\= \prod_{(i,j) \in A} t^{i}
  q^{-1} (1 - q^{\mathrm{Arm}_{A}(i, j)} t^{\mathrm{Leg}_A(i, j) +
    1})^{-1} \prod_{(i,j) \in B} q^{1 + j + u} t^{N - 1}
    (q^{\mathrm{Arm}_B(i, j)} t^{\mathrm{Leg}_B (i, j)+ 1} - 1)^{-1}
      \times\\\times \frac{ G^{(q,t)}_{AB}(q^{-u} t^{-N}) G^{(q,t)}_{A
          \varnothing}(q^{1+ v} t^{N - 1}) G^{(q,t)}_{B
          \varnothing}(t^N) } { G^{(q,t)}_{A\varnothing}(q^{-u})
        G^{(q,t)}_{B \varnothing}(q^{u + v + 2} t^{2 N - 2})},
\end{multline}
One can also transform the averages of positive power sums $p_n$ to
negative ones $p_{-n}$ and vice versa:
\begin{equation}
  \label{eq:59}
  \langle f(x_i) \rangle_{u,v,N,q,t} = \langle f(q^{1-v}
  x_i^{-1})\rangle_{-u - v - 2 + 2 \beta - 2 \beta N,v,N,q,t}
\end{equation}
which leads to
\begin{multline}
  \left\langle M_A \left(p_n + \frac{q^n/t^n - q^{-n v}}{1 - t^{-n}}
    \right) M_B ( p_{-n}) \right\rangle_{u,v,N,q,t} =\\
  =q^{(|A|-|B|)(1-v)} \left\langle M_A \left( p_{-n} + \frac{t^n/q^n -
        q^{n v}}{1 - t^n} \right) M_B (p_n)
  \right\rangle_{-u-v-2+2\beta - 2\beta N,v,N,q,t}.
\end{multline}

We remind the result from~\cite{genMD}:
\begin{multline}
  \label{eq:57}
  \left \langle M^{(q,t)}_{AB} \left( q^{u+v+1} t^{2N-1} \left| p_n ,
        -p_n - \frac{1 - q^{-n v}}{1 - t^{-n}} \right.\right) \right
  \rangle_{u,v,N,q,t} =\\
  = (-1)^{|A|} q^{-(v+1)|A|-(u+2v+3)|B| + \sum_{(i,j)\in A} j + 2
    \sum_{(i,j)\in B}j} t^{|C|-(2N+3)|B| -\sum_{(i,j)\in B} i}
  \times \\
  \times \frac{f_A^{+} \left( t^{-N} \right) f_A^{+} \left( q^{-u-1}
      t^{1-N} \right) f_B^{+} \left( q^{u+v+1}t^{N-1} \right) f_B^{+}
    \left( q^vt^N \right)}{C'_A C'_B G_{BA}^{(q,t)} \left( q^{-u-v-1}
      t^{1-2N} \right)}.
\end{multline}
We also give an alternative average of generalized Macdonald
polynomial (notice the difference in shifts of the power sums $p_n$)
\begin{multline}
  \label{eq:63}
  \left \langle M_{AB}^{(q,t)}\left( q^{-u-1} t \left| p_{-n} +
        \frac{(t/q)^n - q^{nv}}{1-t^n} , -p_{-n} - \left( \frac{t}{q}
        \right)^n \frac{1 - (t/q)^n}{1-t^n} \right. \right) \right
  \rangle_{u,v,N,q,t} =\\
  =(-1)^{|A|} q^{-2|B|+u|A|} t^{|B|-|A|} t^{\sum_{(i,j)\in B} i + 2
    \sum_{(i,j)\in A} i} q^{- \sum_{(i,j)\in A} j} \times\\
  \times \frac{f_A^{-}\left( t^N q^u \right) f_A^{-}\left( t^{1-N}
      q^{-v-1} \right) f_B^{-}\left( t^{N+1} q^{-1} \right)
    f_B^{-}\left( t^{2-N} q^{-u-v-2} \right)}{ C'_A(q,t) C'_B(q,t)
    G_{BA}^{(q,t)} \left( q^{u+1} t^{-1} \right)}
\end{multline}

Finally, we were able to find the most general factorized formula for
the average of \emph{two generalized} Macdonald polynomials (or
\emph{generalized bifundamental kernel}
$\widetilde{N}^{(q,t)}_{AB,CD}(u,v,N|x)$), which gives all the
averages above as special cases\footnote{We have checked this formula
  up to the third level.}:
\begin{multline}
  \label{eq:58}
  \langle \widetilde{N}^{(q,t)}_{AB,CD}(u,v,N|x) \rangle_{u,v,N,q,t} =
  \Biggl\langle \sum_{E,F} \left( \frac{t}{q} \right)^{|E|+|F|}
  \frac{1}{||M_E^{(q,t)}||^2 ||M_F^{(q,t)}||^2 } \times \\
  \times M^{*(q,t)}_{AB/EF} \left( q^{u+1} t^{-1} \left| - \left(
        \frac{t}{q} \right)^n p_{-n} - \frac{1 - (t/q)^n}{1 - t^n}
      q^{nv} , p_{-n} + \frac{(t/q)^n -
        q^{n v}}{1 - t^n} \right. \right)\times \\
  \times M^{(q,t)}_{CD/EF} \left( q^{u+v+1} t^{2N-1} \left| p_n , -p_n
      - \frac{1 - q^{-n v}}{1 - t^{-n}} \right.\right)
  \Biggr\rangle_{u,v,N,q,t} = \\
  = (-1)^{|B|+|C|} q^{-2 |A| - (v+1) |C| + u |B| - (u+2v+3)|D|}
  t^{|A|- |B| + |C| -
    (2N + 3)|D|}\times\\
  \times t^{\sum_{(i,j)\in A} i + 2\sum_{(i,j)\in B} i -
    \sum_{(i,j)\in D} i} q^{ - \sum_{(i,j)\in B} j +
    \sum_{(i,j)\in C} j + 2\sum_{(i,j)\in D} j} \times \\
  \times \frac{z_{\mathrm{bifund}}^{(q,t)}\left([A,B],[C,D],\frac{u+1
        - \beta}{2} , \frac{u+v + 2 \beta N - \beta +1 }{2},
      -\frac{v}{2}-1+ \beta \right)}{C'_A(q,t) C'_B(q,t) C'_C(q,t)
    C'_D(q,t) G_{AB}^{(q,t)} (q^{u+1}t^{-1}) G_{DC}^{(q,t)}
    (q^{-u-v-1}t^{1-2N}) },
\end{multline}
where $C'_A(q,t) = \prod_{(i,j) \in A} (1 - q^{A_i - j}
t^{A^{\mathrm{T}}_j - i + 1}) $.

Again, one can use the symmetry~\eqref{eq:59} to write
Eq.~\eqref{eq:58} in an alternative form:
\begin{multline}
  \label{eq:60}
  \langle \widetilde{N}^{(q,t)}_{AB,CD}(u,v,N|x)
  \rangle_{-u-v-2+2\beta - 2\beta N,v,N,q,t} = \langle
  \widetilde{K}^{(q,t)}_{AB,CD}(u,v,N|q^{1-v}x^{-1})
  \rangle_{u,v,N,q,t}  =\\
  = q^{(v-1)(|A|+|B|-|C|-|D|)} \Biggl\langle \sum_{E,F} \left(
    \frac{t}{q} \right)^{|E|+|F|}
  \frac{1}{||M_E^{(q,t)}||^2 ||M_F^{(q,t)}||^2 }\times\\
  \times M^{*(q,t)}_{AB/EF} \left( q^{-u-v-1} t^{1-2N} \left| - \left(
        \frac{t}{q} \right)^n p_n - \frac{1 - (q/t)^n}{1 - t^{-n}},
      p_n + \frac{(q/t)^n -
        q^{-n v}}{1 - t^{-n}} \right. \right)\times \\
  \times M^{(q,t)}_{CD/EF} \left( q^{-u-1} t \left| p_{-n} , -p_{-n}
      -\left( \frac{t}{q} \right)^n \frac{1 - q^{n v}}{1 - t^n}
    \right. \right) \Biggr\rangle_{u,v,N,q,t}.
\end{multline}

\section{Open topological string amplitude on resolved conifold}
\label{sec:open-topol-string}
\begin{figure}[h]
  \centering
  \begin{equation}
    Z \left( \left.
  \begin{smallmatrix}
     & P & \\
     A&  & B\\
     & R &
  \end{smallmatrix}
\right| Q, q,t \right) \quad = \quad \parbox{2.8cm}{\includegraphics[width=2.8cm]{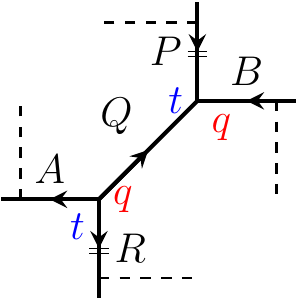}}\notag
  \end{equation}
  \caption{Resolved conifold with stacks of Lagrangian A-branes on
    each leg of the toric diagram. Refined open string amplitude
    depends on four Young diagrams $A$, $B$, $R$ and $P$ and the
    K\"ahler parameter $Q$ of the conifold.}
  \label{fig:6}
\end{figure}

In this appendix we write down the basic building block of the toric
diagrams related to $5d$ quiver gauge theories. It is given by an open
refined topological string amplitude in the resolved conifold
background depicted in Fig.~\ref{fig:6}.

Using the IKV refined topological vertex one gets the following answer
for this amplitude:
\begin{multline}
  \label{eq:47}
  Z \left( \left.
  \begin{smallmatrix}
     & P & \\
     A&  & B\\
     & R &
  \end{smallmatrix}
\right| Q, q,t \right) = \sum_C (-Q)^{|C|} C_{ACR}(t,q)
C_{B^{\mathrm{T}} C^{\mathrm{T}} P^{\mathrm{T}}}(q,t) =\\
= Z \left( \left.
  \begin{smallmatrix}
     & \varnothing & \\
     \varnothing&  & \varnothing\\
     & \varnothing &
  \end{smallmatrix}
\right| Q, q,t \right)  q^{\frac{||R||^2 - ||P||^2}{2}}
t^{\frac{||P^{\mathrm{T}}||^2 - ||R^{\mathrm{T}}||^2}{2}} \left( \frac{q}{t}
\right)^{\frac{|A| - |B|}{2}}
M_R^{(q,t)}(t^{-\rho}) M_{P^{\mathrm{T}}}^{(t,q)} (q^{-\rho})
G_{RP}^{(q,t)} \left( \sqrt{\frac{q}{t}} Q \right) \times\\
\times \sum_C (-Q)^{|C|}
\chi_{A^{\mathrm{T}}/C^{\mathrm{T}}} \left(p_n (t^{-\rho} q^{-R}) - p_n \left(
    \sqrt{\frac{q}{t}} Q t^{-\rho} q^{-P}  \right)\right) \chi_{B/C}
\left( p_n (q^{-\rho} t^{-P^{\mathrm{T}}}) - p_n \left(
    \sqrt{\frac{t}{q}} Q q^{-\rho} t^{-R^{\mathrm{T}}} \right) \right)
\end{multline}
where $Z \left( \left.
  \begin{smallmatrix}
     & \varnothing & \\
     \varnothing&  & \varnothing\\
     & \varnothing &
  \end{smallmatrix}
\right| Q, q,t \right) = \prod_{i, j \geq 1} \left(1 - Q
q^{i-\frac{1}{2}} t^{j - \frac{1}{2}}\right)$ is the closed refined
string amplitude on the conifold.

One can perform a flop transformation on the conifold geometry. We
employ the following symmetry of the closed string amplitude:
\begin{equation}
  \label{eq:51}
  Z \left( \left.
  \begin{smallmatrix}
     & \varnothing & \\
     \varnothing&  & \varnothing\\
     & \varnothing &
  \end{smallmatrix}
\right| Q, q,t \right) = Z \left( \left.
  \begin{smallmatrix}
     & \varnothing & \\
     \varnothing&  & \varnothing\\
     & \varnothing &
  \end{smallmatrix}
\right| Q, q^{-1} ,t^{-1} \right) =  \left(- Q^{-1} \sqrt{q^{-1} t^{-1}}\right)^{\frac{1}{12}} Z \left( \left.
  \begin{smallmatrix}
     & \varnothing & \\
     \varnothing&  & \varnothing\\
     & \varnothing &
  \end{smallmatrix}
\right| Q^{-1} , q,t \right).
\end{equation}

The flopped open string amplitude is related to the original amplitude
with K\"ahler parameter reversed:
\begin{multline}
  \label{eq:48}
  Z_{\mathrm{flopped}}\left( \left.
  \begin{smallmatrix}
     & R & \\
     A&  & B\\
     & P &
  \end{smallmatrix}
\right| Q, q,t \right) = \left(- Q^{-1} \sqrt{q^{-1}
  t^{-1}}\right)^{\frac{1}{12}} q^{\frac{||R||^2 - ||P||^2 +
  ||A^{\mathrm{T}}||^2 - ||B^{\mathrm{T}}||^2}{2}}
t^{-\frac{||R^{\mathrm{T}}||^2 -
    ||P^{\mathrm{T}}||^2 + ||A||^2 - ||B||^2}{2}} \times\\
\times \left( \frac{q}{t} \right)^{\frac{|A|-|B|}{2}}
(-Q)^{|R|+|P|+|A|+|B|} Z\left( \left.
  \begin{smallmatrix}
     & P & \\
     A&  & B\\
     & R &
  \end{smallmatrix}
\right| Q^{-1} , q,t \right).
\end{multline}
The multipliers in Eq.~\eqref{eq:48} combine with the change of
framing in the adjacent edges induced by the flop. The answer for
any closed string amplitude, which includes the flopped part is given
simply by
\begin{equation}
  \label{eq:49}
  Z_{\mathrm{flopped}} (Q,Q_{\mathrm{adjacent}}, Q_i) = \left(- Q^{-1} \sqrt{q^{-1}
  t^{-1}}\right)^{\frac{1}{12}} Z (Q^{-1}, Q Q_{\mathrm{adjacent}}, Q_i),
\end{equation}
where in the right hand side the original K\"ahler parameter of the
conifold is reversed and the K\"ahler parameters of the two-cycles
adjacent to the flopped conifold are shifted by $Q$.

\section{Useful identities}
\label{sec:useful-identities}
One has the following identity for the power sum symmetric functions
$p_n(x) = \sum_{i \geq 1} x_i^n$:
\begin{equation}
  \label{eq:23}
  p_n(q^Y t^{\rho}) = - \left( \frac{t}{q} \right)^{n/2} \frac{1 -
    q^n}{1 - t^n} p_n (t^{-Y^{\mathrm{T}}} q^{-\rho})
\end{equation}
where $\rho_i = \frac{1}{2} - i$ and $Y$ is a Young diagram.

Macdonald polynomials satisfy the following ``transposition''
identities:
\begin{equation}
  \label{eq:24}
  M_Y^{(t,q)} (p_n) = (-1)^{|Y|} h_{Y^{\mathrm{T}}}(q,t) M_{Y^{\mathrm{T}}}^{(q,t)}\left( -
    \frac{1 - q^n}{1 - t^n} p_n \right),
\end{equation}
where
\begin{gather}
  \label{eq:25}
  h_{Y^{\mathrm{T}}}(q,t) = \frac{C'_{Y^{\mathrm{T}}} (q,t)}{C'_Y (t,q)}
  = \prod_{(i,j) \in Y} \frac{1 - t^{Y_i - j + 1} q^{Y^{\mathrm{T}}_j
      - i}}{1 - t^{Y_i - j}
    q^{Y^{\mathrm{T}}_j - i + 1}},\\
      C'_Y (q,t) = \prod_{(i,j) \in Y} \left( 1 - q^{Y_i - j}
        t^{Y^{\mathrm{T}}_j - i + 1} \right).\notag
\end{gather}
Combining Eqs.~\eqref{eq:23} and~\eqref{eq:25} we get the identity,
which will be useful in refined topological string computations:
\begin{equation}
  \label{eq:14}
  M_{\alpha}^{(t,q)} \left( p_n (q^{\beta} t^{\rho}) - p_n(Q t^{\rho}) \right)
  = (-1)^{|\alpha|} h_{\alpha^{\mathrm{T}}}(q,t) M_{\alpha^{\mathrm{T}}}^{(q,t)} \left( p_n (t^{-\beta^{\mathrm{T}}} q^{-\rho}) - p_n(Q q^{-\rho}) \right).
\end{equation}

The following identity involving Nekrasov functions is also useful:
\begin{equation}
  \label{eq:38}
  \prod_{i=1}^{N_1} \prod_{j=1}^{N_2} \frac{1 - Q q^{j - W_i - \frac12} t^{i -
      Y^{\mathrm{T}}_j - \frac12}}{1 - Q q^{j-\frac12} t^{i-\frac12}} = \frac{G^{(q,t)}_{Y
      W}(q^{\frac12} t^{-\frac12} Q)}{G^{(q,t)}_{Y \varnothing}(Q
    q^{\frac12} t^{N_1 - \frac12})
    G^{(q,t)}_{\varnothing W}(Q q^{N_2})},
\end{equation}
and in particular for $N_{1,2}\to \infty$ (we assume $|q|, |t| < 1$):
\begin{equation}
  \label{eq:31}
  \prod_{i,j\geq 1} \frac{1 - Q q^{j - W_i - \frac12} t^{i -
      Y^{\mathrm{T}}_j - \frac12}}{1 - Q q^{j-\frac12} t^{i-\frac12}} = G^{(q,t)}_{YW}\left(\sqrt{\frac{q}{t}} Q\right).
\end{equation}
One can exchange the diagrams in Nekrasov function using the identity
\begin{equation}
  \label{eq:44}
  G_{BA}^{(q,t)}\left( \sqrt{\frac{q}{t}} Q \right) = (-Q)^{|A| + |B|}
  q^{\frac{||B||^2 - ||A||^2}{2}}  t^{\frac{||A^{\mathrm{T}}||^2 - ||B^{\mathrm{T}}||}{2}}G_{AB}^{(q,t)}\left( \sqrt{\frac{q}{t}} Q^{-1} \right),
\end{equation}
where $||R||^2 = \sum_i R_i^2$.

When one of the diagrams is zero, there is a nice expression in terms
of Macdonald polynomials:
\begin{gather}
  \label{eq:45}
  G_{R \varnothing}^{(q,t)}\left(\sqrt{\frac{q}{t}} Q\right) =
  \prod_{(i,j) \in R} (1 - Q q^{j-\frac{1}{2}} t^{\frac{1}{2} -i}) =
  \frac{M_R^{(q,t)} \left( \frac{1 - Q^n q^{\frac{n}{2}}
        t^{-\frac{n}{2}}}{t^{- \frac{n}{2}} - t^{\frac{n}{2}}}
    \right)}{M_R^{(q,t)} \left( \frac{1}{t^{- \frac{n}{2}} -
        t^{\frac{n}{2}}} \right)} = \frac{M_R^{(q,t)} \left(
      p_n(t^{-\rho}) - p_n \left( Q \sqrt{\frac{q}{t}} t^{-\rho}
      \right)\right)}{M_R^{(q,t)} \left(t^{-\rho} \right)},\\
  G_{\varnothing P}^{(q,t)}\left(\sqrt{\frac{q}{t}} Q \right) =
  \prod_{(i,j) \in P} (1 - Q q^{\frac{1}{2}-j} t^{i-\frac{1}{2}}) =
  \frac{M_{P^{\mathrm{T}}}^{(t,q)} \left( \frac{1 - Q^n t^{\frac{n}{2}}
        q^{-\frac{n}{2}}}{q^{- \frac{n}{2}} - q^{\frac{n}{2}}}
    \right)}{M_{P^{\mathrm{T}}}^{(t,q)} \left( \frac{1}{q^{- \frac{n}{2}} -
        q^{\frac{n}{2}}} \right)} = \frac{M_{P^{\mathrm{T}}}^{(t,q)} \left(
      p_n(q^{-\rho}) - p_n \left( Q \sqrt{\frac{t}{q}} q^{-\rho}
      \right)\right)}{M_{P^{\mathrm{T}}}^{(t,q)} \left(q^{-\rho} \right)},
  \label{eq:46}
\end{gather}
and also
\begin{equation}
  \label{eq:77}
   M_R^{(q,t)} \left(t^{-\rho} \right) = \frac{t^{\frac{||R^{\mathrm{T}}||^2}{2}}}{C_R'(q,t)}.
\end{equation}

\end{document}